\documentclass[aps,prd,floats,twocolumn,epsf,nofootinbib,showpacs]{revtex4}
\usepackage{latexsym}
\usepackage{amssymb}
\usepackage[dvips]{graphicx}
\usepackage[applemac]{inputenc}
\usepackage[nottoc]{tocbibind} %prevents the entry "Table of
\usepackage{float}
\usepackage{fancyhdr}
\usepackage{amsfonts}
\usepackage{amsmath}
\usepackage{color}
\usepackage{mathrsfs}
\usepackage{bm}% bold math
\usepackage{epsfig}% bold math

\newcommand{\be}{\begin{equation}}
\newcommand{\ee}{\end{equation}}

\def\lsim{\mathrel{\raise.3ex\hbox{$<$\kern-.75em\lower1ex\hbox{$\sim$}}}}

\def\gsim{\mathrel{\raise.3ex\hbox{$>$\kern-.75em\lower1ex\hbox{$\sim$}}}}

\def\beq{\begin{eqnarray}}
\def\eeq{\end{eqnarray}}
\def\bea{\begin{eqnarray}}
\def\eea{\end{eqnarray}}

\begin{document}

\title{Implications of CoGeNT's New Results For Dark Matter}

\author{Dan Hooper$^{a,b}$ and Chris Kelso$^{c}$}

\address{
$^a$Center for Particle Astrophysics, Fermi National Accelerator
Laboratory, Batavia, IL 60510 \\ 
$^b$Department of Astronomy and Astrophysics, University of Chicago,
Chicago, IL 60637 \\
$^c$Department of Physics, University of Chicago, Chicago, IL 60637}

\begin{abstract}

The CoGeNT collaboration has recently presented the results of their first 15 months of data, including the measurement of the spectrum of nuclear recoil candidate events, and the time variation of those events. These results appear consistent with the signal anticipated from a relatively light dark matter particle scattering elastically with nuclei. In this paper, we independently analyze the data set collected by CoGeNT and explore the implications of these results for dark matter. We find that the observed spectrum and rate is consistent with originating from dark matter particles with a mass in the range of 4.5-12 GeV and an elastic scattering cross section with nucleons of approximately $\sim$$10^{-40}$ cm$^2$. We confirm the conclusion of the CoGeNT collaboration that the data also includes a somewhat statistically significant (2.7$\sigma$) indication of annual modulation, with a phase, period, and amplitude consistent with that predicted for dark matter. CoGeNT's phase is also consistent with the annual modulation reported by the DAMA/LIBRA collaboration. We also discuss the null results reported by CDMS and XENON100, and comment on the prospects for other experiments to detect a dark matter particle with the properties implied by CoGeNT.

\end{abstract}

\pacs{95.36.+x; FERMILAB-PUB-11-248-A}

\maketitle

%%%%%%%%%%%%%%%%%%%%%%%%%%%%%%%%%%%%%%%%%%%%%%%%%%%%%%%%%%%%%%%%%%%%%%

\section{Introduction}

Although there exists an abundance of evidence that the vast majority of matter in our universe is non-baryonic and does not significantly emit, reflect, or absorb light, the nature of this dark matter remains unknown. Among the techniques being pursued to identify the particle identity of dark matter are direct detection experiments, which are designed to observe particles of dark matter in the Galactic Halo through their elastic scattering with nuclei in a target material.

Many of the technologies and target materials currently used in direct detection experiments are most sensitive to dark matter particles with masses greater than $\sim$$10$ GeV. By virtue of their very low electronic noise, however, the P-type point contact germanium detectors employed by the CoGeNT collaboration are able to detect very low energy scattering events and thus, despite their modest target mass of 475 grams (330 grams fiducial), are quite sensitive to low mass WIMPs.

In early 2010, the CoGeNT collaboration reported the observation of $\sim$100 events above expected backgrounds over a period of 56 days, with ionization energies in the range of approximately 0.4 to 1.0 keV~\cite{cogent}. One possible interpretation of these events is the elastic scattering of dark matter particles with a mass in the range of approximately 5-10 GeV and a cross section with nucleons on the order of $\sim$$10^{-40}$ cm$^2$~\cite{zurek}.

Very recently, the CoGeNT collaboration has announced the results of their analysis of a full 15 months of data~\cite{newcogent}. This larger data set has been used to provide a much more detailed measurement of the spectrum of observed events. Furthermore, their analysis has revealed a time variation in the rate of low energy nuclear recoil events, with a quoted significance of 2.8$\sigma$. As a result of the Earth's motion around the Sun and relative to the rest frame of the dark matter halo, the rate of dark matter elastic scattering events is predicted to vary with an annual cycle~\cite{modulation}. The modulation reported by the CoGeNT collaboration is consistent in amplitude, phase, and period with that predicted to arise from elastically scattering dark matter~\cite{kelso}.  

The only other direct detection experiment to report the observation of an annual modulation in their event rate is DAMA/LIBRA. In particular, the DAMA/LIBRA collaboration reports a high significance (8.9$\sigma$) detection of annual modulation with a phase and period consistent with elastically scattering dark matter~\cite{damanew}. The spectrum of the signals reported by DAMA/LIBRA and CoGeNT each point toward a similar range of dark matter parameter space~\cite{consistent}. Furthermore, the range of dark matter mass implied by CoGeNT and DAMA/LIBRA is very similar to that required to explain the spectrum of gamma rays observed by the Fermi Gamma Ray Space Telescope (FGST) from the the inner 0.5$^{\circ}$ around the Galactic Center~\cite{Hooper:2010mq}, and for the observed synchrotron emission known as the WMAP Haze~\cite{Hooper:2010im}.

%\begin{twocolumn}
\begin{figure*}[t]
\centering
{\includegraphics[angle=0.0,width=3.4in]{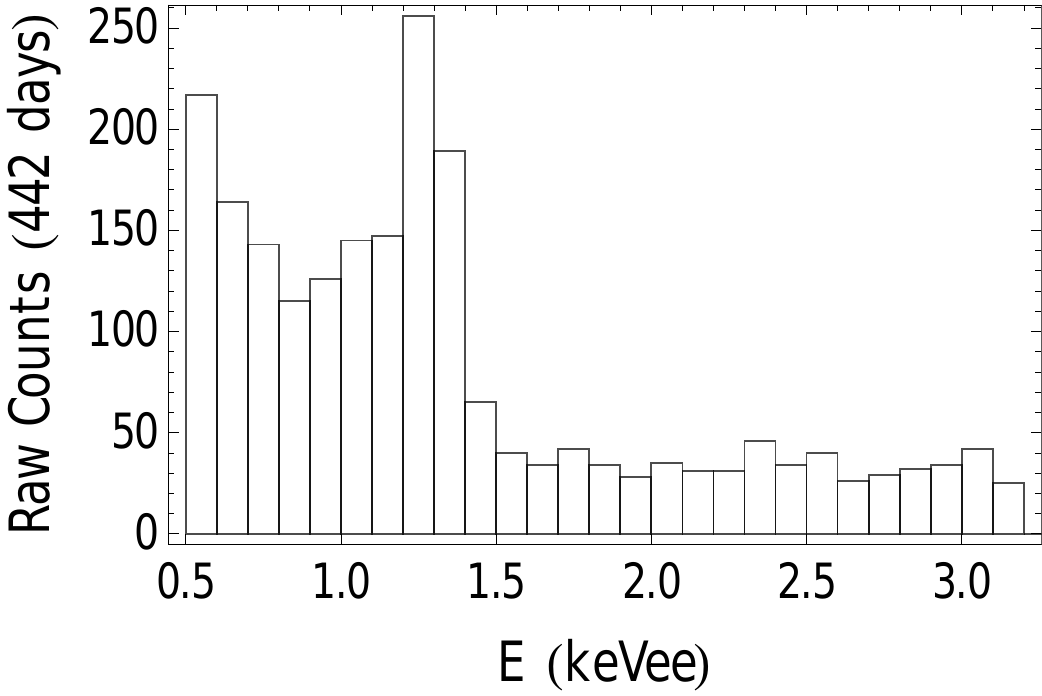}}
{\includegraphics[angle=0.0,width=3.4in]{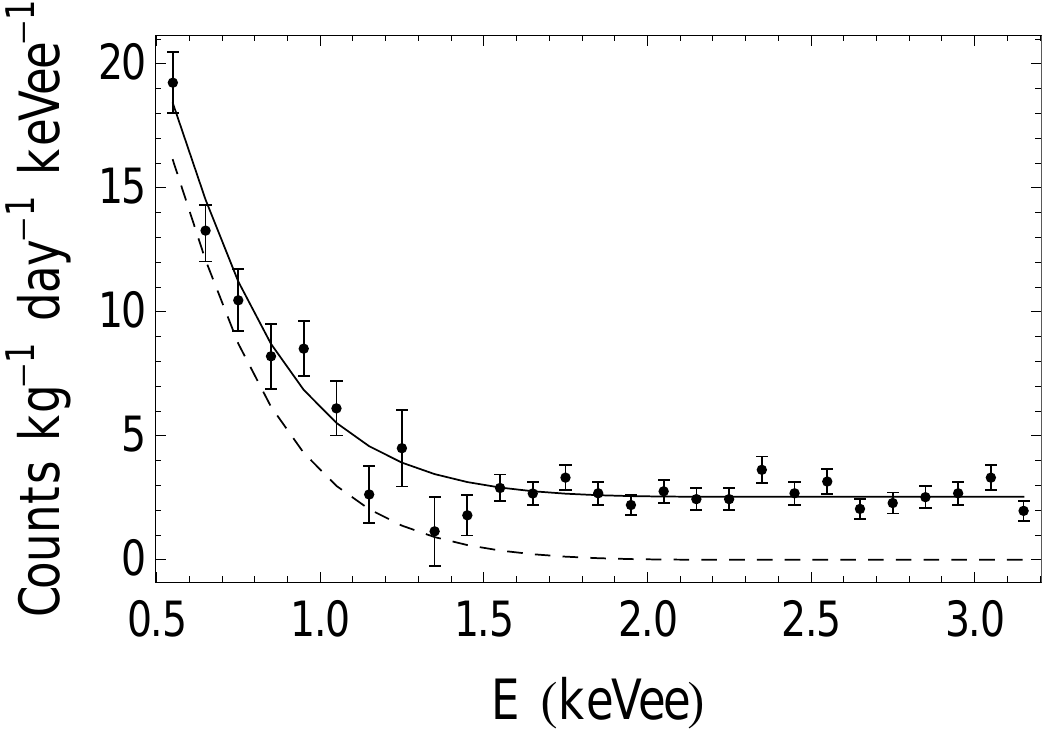}}
\caption{In the left frame, we show the raw spectrum of events reported by CoGeNT. In the right frame, the spectrum is shown after subtracting the predicted contribution from L-shell electron capture peaks and correcting for the detector efficiency. For comparison, we show the spectrum predicted for a dark matter particle with a mass of 7 GeV, an elastic scattering cross section with nucleons of $1.2\times 10^{-40}$ cm$^2$, a local density of 0.3 GeV/cm$^3$, and with a velocity distribution described by $v_0=250$ km/s and $v_{\rm esc}=550$ km/s.  The dashed line represents the spectrum of dark matter events alone, while the solid line is the dark matter spectrum plus a flat background.}
\label{spec}
\end{figure*}
%\end{twocolumn}

In this paper, we present an independent analysis of the data recently provided by CoGeNT and discuss the implications of this data for particle dark matter. In Sec.~\ref{spectrum}, we calculate the spectrum of events predicted to result from elastically scattering dark matter particles and compare this prediction to that reported by CoGeNT. In doing so, we find that for a reasonable range of dark matter velocity distributions, the spectrum of the CoGeNT excess is consistent with a dark matter particle with a mass in the range of 4.5 to 12 GeV. In Sec.~\ref{annual}, we discuss the properties of CoGeNT's annual modulation. In Sec.~\ref{others}, we consider the constraints on a dark matter interpretation of the CoGeNT signal from other direct detection experiments and discuss the implications of the CoGeNT result for other dark matter searches currently being conducted. In Sec.~\ref{summary} we summarize our results and conclusions.

%%%

%%%%%%%%%%%%%

\section{CoGeNT's Spectrum and Elastically Scattering Dark Matter}
\label{spectrum}

\begin{table*}[ht]
\begin{tabular}{|c|c|c|c|c|c|}
\hline
Isotope  & Total Events & Uncertainty (\%) & Energy (keV) & Half-Life (days)  \\
\hline
As$^{73}$  & 12.74 & 33.48 & 1.414 & 80  \\
\hline
Ge$^{68}$  &    639.0 & 1.35  &   1.298   &     271 \\
\hline 
Ga$^{68}$  &    52.83 & 5.11  &   1.194   &   271 \\ 
\hline 
Zn$^{65}$  &    211.2 & 2.23  &   1.096   &   244 \\ 
\hline 
Ni$^{56}$  &    1.53 & 23.46  &  0.926      &   5.9 \\ 
\hline 
Co$^{56,58}$&   9.44 & 44.9&   0.846     &   71 \\ 
\hline 
Co$^{57}$   &  2.59  &8.0&   0.846   &   271 \\ 
\hline 
Fe$^{55}$   &   44.94 & 11.63  &  0.769      &   996 \\ 
\hline 
Mn$^{54}$   &   21.09 & 9.34  &   0.695   &   312 \\ 
\hline 
Cr$^{51}$   &   2.94 & 15.29  &   0.628    &   28 \\ 
\hline 
V$^{49}$    &   14.91 & 12.26  &   0.564    &   330 \\ 
\hline
\end{tabular}
\caption{Characteristics of the backgrounds from cosmogenically-activated radioisotopes decaying via electron capture (EC). By measuring the corresponding K-shell peaks, the properties of these L-shell peaks can be well constrained. Listed here are the total number of events predicted in each L-shell peak (over all time after the beginning of the current CoGeNT data set), along with the fractional uncertainty in this quantity. The half-life of each decay is also given, along with the central energy of each peak.}
\end{table*}

The CoGeNT detector, located in Northern Minnesota's Soudan Underground Laboratory, observes nuclear recoil events as ionization. In the left frame of Fig.~\ref{spec}, we show the raw spectrum of events (between 0.5 and 3.2 keVee) observed by CoGeNT as a function of ionization energy (in keV-electron equivalent, keVee). Some of these events are the result of cosmogenically-activated radioisotopes decaying via electron capture. Most apparent are peaks appearing near 1.1 and 1.3 keVee, which result from Zn$^{65}$ and Ge$^{68}$, respectively. These backgrounds correspond to the L-shell peaks associated with the isotopes listed in Table~1. By measuring the magnitude of the corresponding K-shell peaks (which appear at higher energies), the rate of these backgrounds can be reliably predicted. In Table~1, the total number of events predicted in each L-shell peak is given (over all time after the beginning of the current CoGeNT data set), along with the fractional uncertainty in this quantity. The half-life of each decay is also listed, along with the central energy of each peak (the width of each peak is determined by the energy resolution of the detector, and varies between 0.0728 and 0.0777 keVee over the relevant energy range).

In the right frame of Fig.~\ref{spec}, we show the spectrum of events reported by CoGeNT, after subtracting the L-shell peaks and correcting for the detector efficiency. Above approximately 1.5 to 2.0 keVee, the spectrum of events observed by CoGeNT is approximately flat and displays no obvious features. At lower energies, however, the rate climbs rapidly. These events appearing below 1.5 keVee are not associated with any known backgrounds~\cite{cogent}.

To assess the hypothesis that the excess events reported by CoGeNT are the product of the elastic scattering of dark matter particles, we will compare CoGeNT's event spectrum to that predicted from dark matter. The spectrum (in nuclear recoil energy) of dark matter induced elastic scattering events is given by~\cite{ls}
\be
\frac{dR}{dE_R} = N_T \frac{\rho_{DM}}{m_{DM}} \int_{|\vec{v}|>v_{\rm
min}} d^3v\, vf(\vec{v},\vec{v_e}) \, \frac{d\sigma}{d E_R},
\label{rate1}
\ee
where $N_T$ is the number of target nuclei, $m_{DM}$ is the mass of
the dark matter particle, $\rho_{DM}$ is the local dark matter
density (which we take to be 0.3 GeV/cm$^3$), $\vec{v}$ is the dark matter velocity in the frame of the
Earth, $\vec{v_e}$ is the velocity of the Earth with respect 
to the galactic halo, and $f(\vec{v},\vec{v_e})$ is the distribution
function of dark matter particle velocities, which we take to be 
the standard Maxwell-Boltzmann distribution:
\be
f(\vec{v},\vec{v_e}) = \frac{1}{(\pi v_0^2)^{3/2}} {\rm
e}^{-(\vec{v}+\vec{v_e})^2/v_0^2}.
\ee
The Earth's speed relative to the galactic halo is given by
$v_e=v_{\odot}+v_{\rm orb}{\rm cos}\,\gamma\, {\rm
cos}[\omega(t-t_0)]$ where $v_{\odot}=v_0+12\,{\rm km/s}$, 
$v_{\rm orb}=30 {\rm km/s}$, ${\rm cos}\,\gamma=0.51$, $t_0$ is the date of the peak in the annual modulation (generally anticipated to lie within several weeks of late May or early June), and $\omega=2\pi/{\rm year}$. We will consider values of $v_0$ over a range of 180 to 320 km/s and values of the galactic escape velocity between 460 and 640 km/s~\cite{velocity}. This function should be thought of as a reasonable, but approximate, parametrization of the dark matter's true velocity distribution. Departures from a Maxwellian velocity distribution are not unexpected and could non-negligibly impact the spectrum of dark matter induced events, as well as the degree of seasonal variation in the rate~\cite{Kuhlen:2009vh,Lisanti:2010qx}.

%Note that the minimum dark matter velocity required to impart a recoil of energy, $E_R$, is given by $v_{\rm min} = \sqrt{E_R m_N/2 \mu^2}$, where $m_N$ is the mass of the target nucleus and $\mu$ is the reduced mass of the dark matter particle and the target nucleus. 

As the germanium isotopes which make up the CoGeNT detector contain little net spin, we consider spin-independent interactions to generate the observed events. In this case, we have
\be
\frac{d\sigma}{d E_R} = \frac{m_N}{2 v^2} \frac{\sigma_n}{\mu_n^2}
\frac{\left[f_p Z+f_n (A-Z)\right]^2}{f_n^2} F^2(q),
\label{cross1}
\ee
where $\mu_n$ is the reduced mass of the dark matter particle and
nucleon (proton or neutron), 
%xxx-liam
%$\sigma_p$
$\sigma_n$
%xxx
 is the scattering cross section of the dark matter 
particle with 
%xxx-liam
%nucleons,
neutrons,
%xxx
$Z$ and $A$ are the atomic and mass numbers of the nucleus, and
$f_{n,p}$ are the coupling strengths of the dark matter particle to
neutrons and protons respectively. Unless stated otherwise, our results have been calculated under the assumption that $f_p=f_n$. The nuclear form factor, $F(q)$, accounts for the finite momentum
transfer in scattering events.  In our calculations, we adopt the
Helm form factor with parameters as described in our previous work~\cite{kelso}. To convert from nuclear recoil energy to the measured ionization energy, we have scaled the results by the quenching factor for germanium as described in Refs.~\cite{Ge,Ge2} ($Q_{\rm Ge}=0.218$ at $E_R=$3 keV, and with the energy dependence predicted by the Lindhard theory~\cite{consistent}).

In the right frame of Fig.~\ref{spec}, we compare the prediction from dark matter to the spectrum of events observed by CoGeNT. In particular, we show the result for the case of a 7 GeV WIMP with an elastic scattering cross section with nucleons of $\sigma_{{\rm DM}-N}=1.2\times 10^{-40}$ cm$^2$, with a velocity distribution described by $v_0=250$ km/s and $v_{\rm esc}=550$ km/s. The dashed line denotes the contribution from dark matter alone, while the solid line also includes a flat background. 

\begin{figure*}[t]
\centering
{\includegraphics[angle=0.0,width=2.49in]{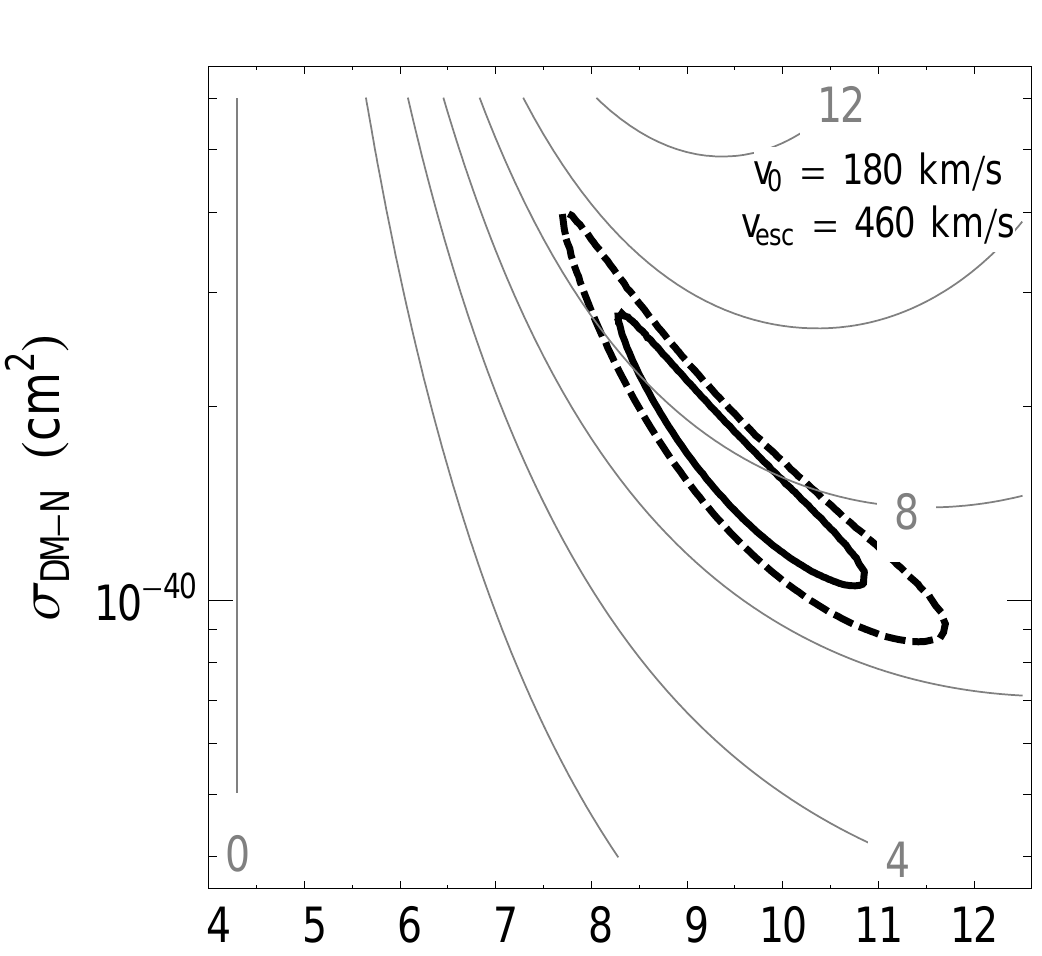}}
{\includegraphics[angle=0.0,width=2.24in]{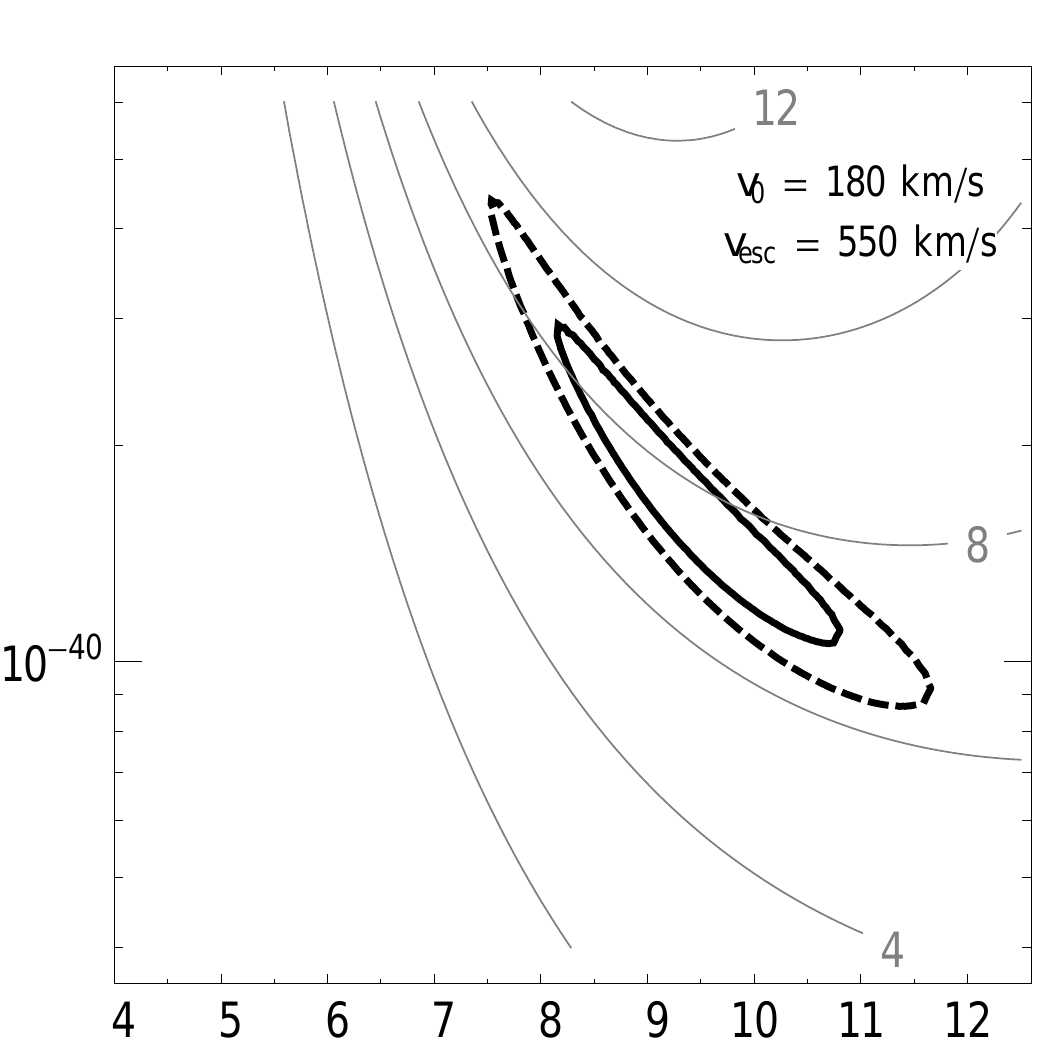}}
{\includegraphics[angle=0.0,width=2.24in]{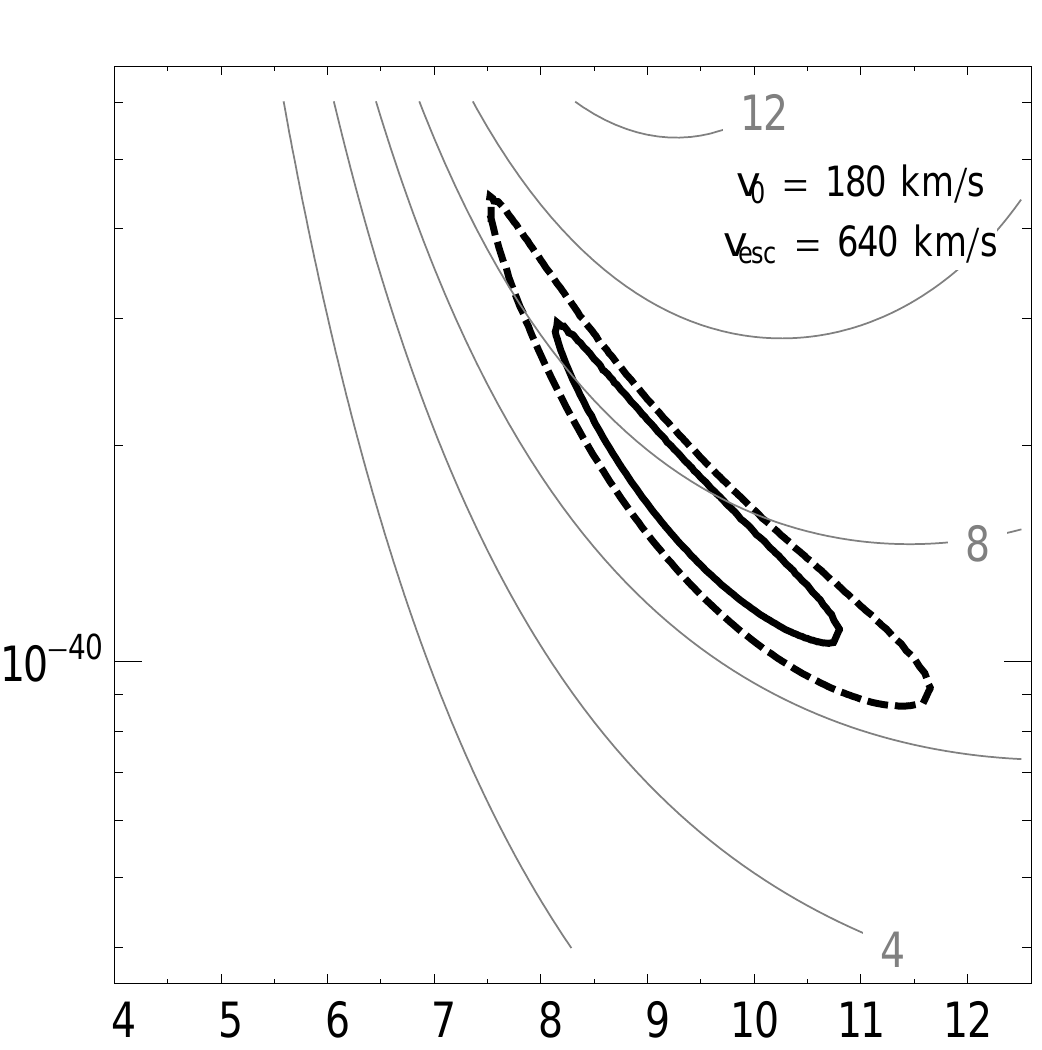}}\\
{\includegraphics[angle=0.0,width=2.49in]{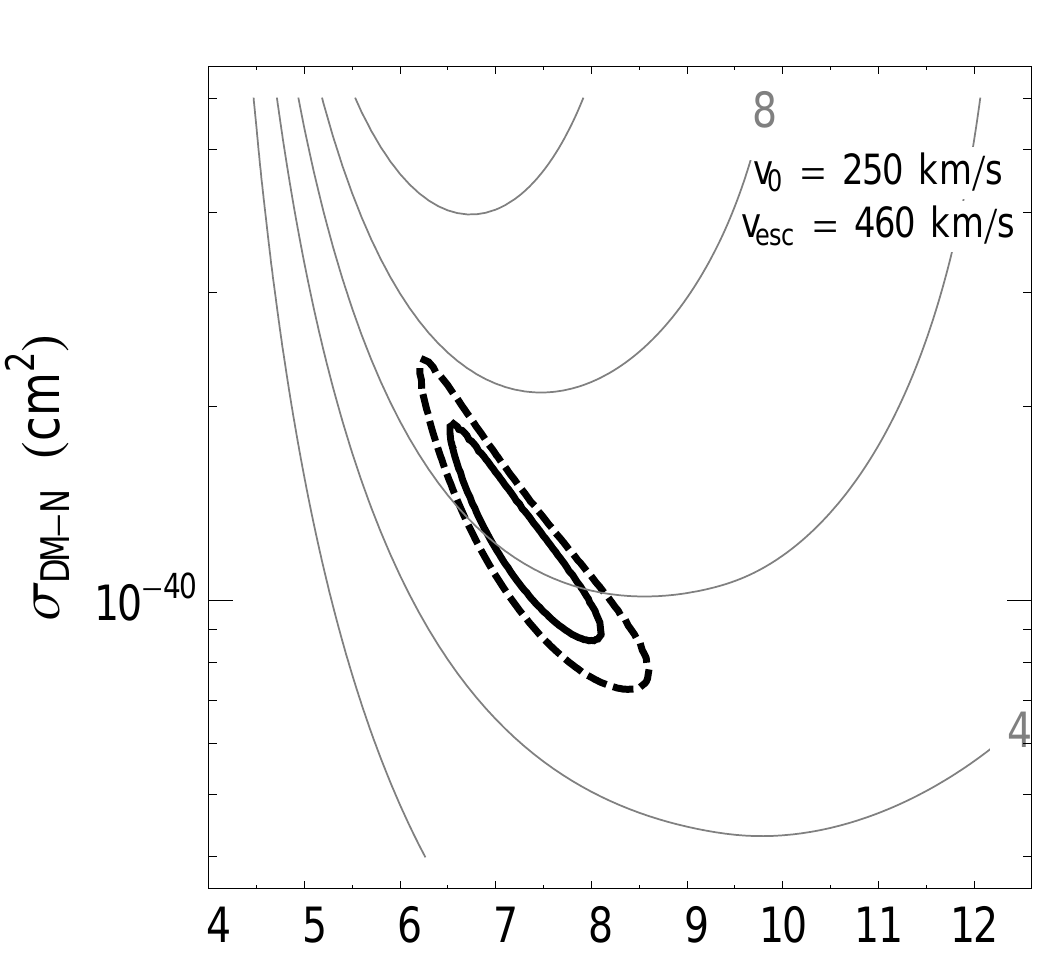}}
{\includegraphics[angle=0.0,width=2.24in]{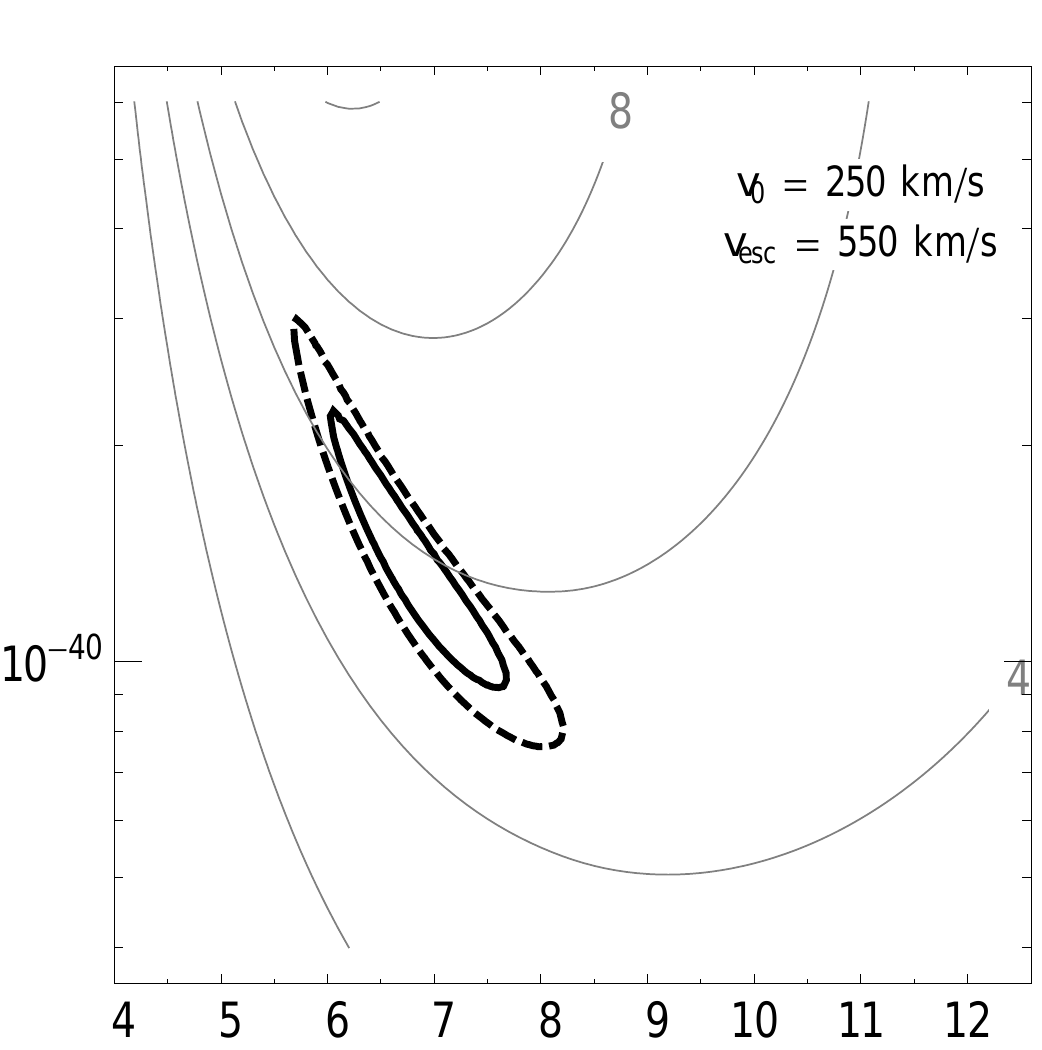}}
{\includegraphics[angle=0.0,width=2.24in]{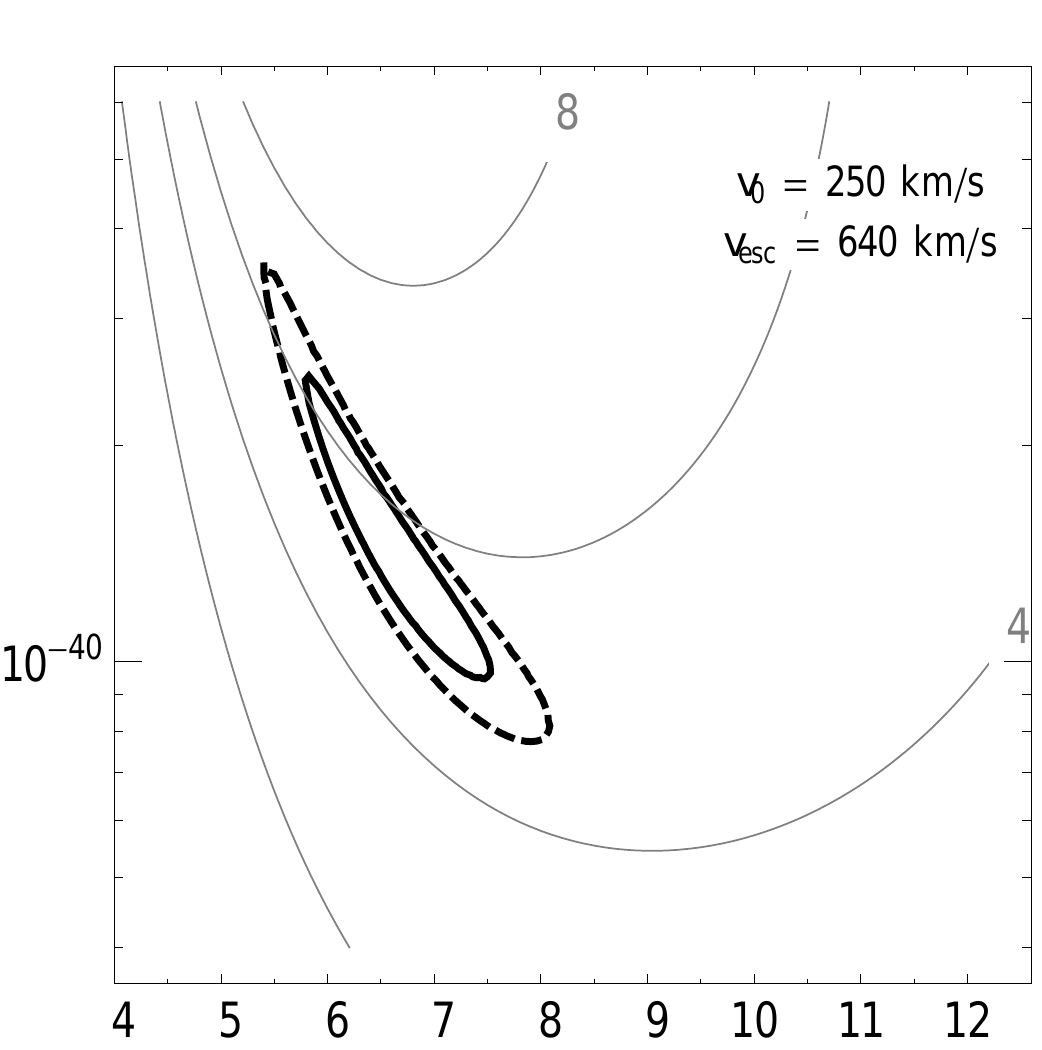}}\\
{\includegraphics[angle=0.0,width=2.49in]{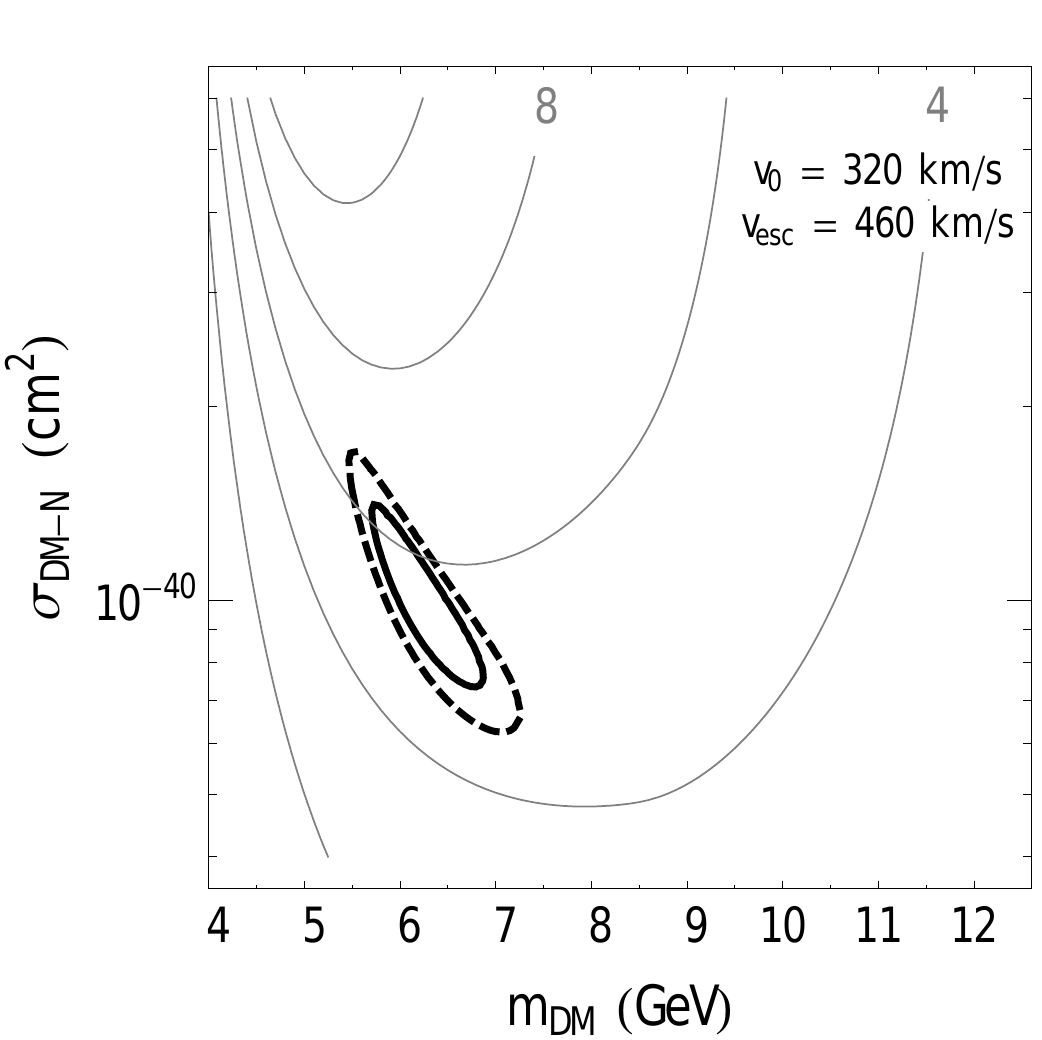}}
{\includegraphics[angle=0.0,width=2.24in]{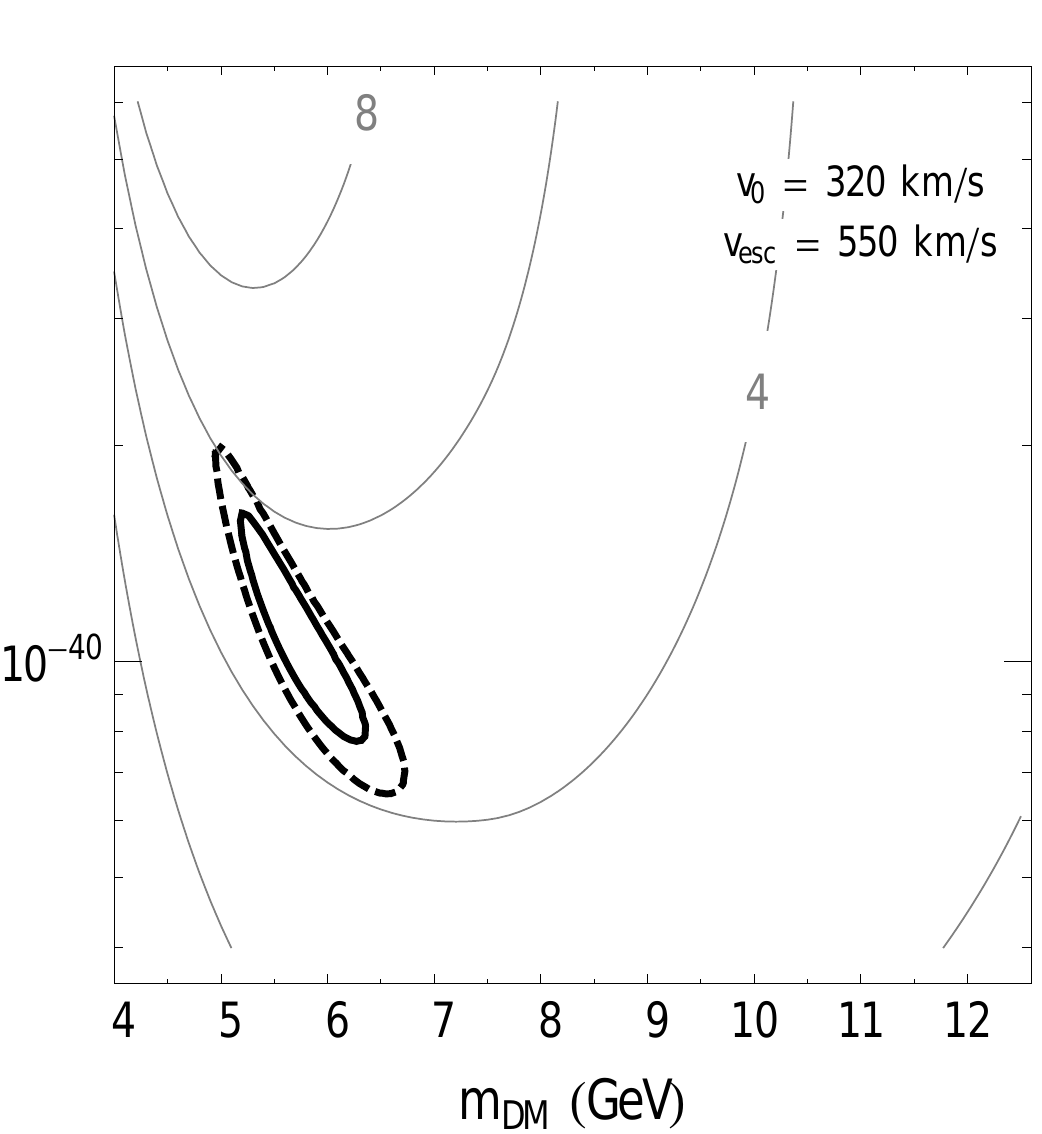}}
{\includegraphics[angle=0.0,width=2.24in]{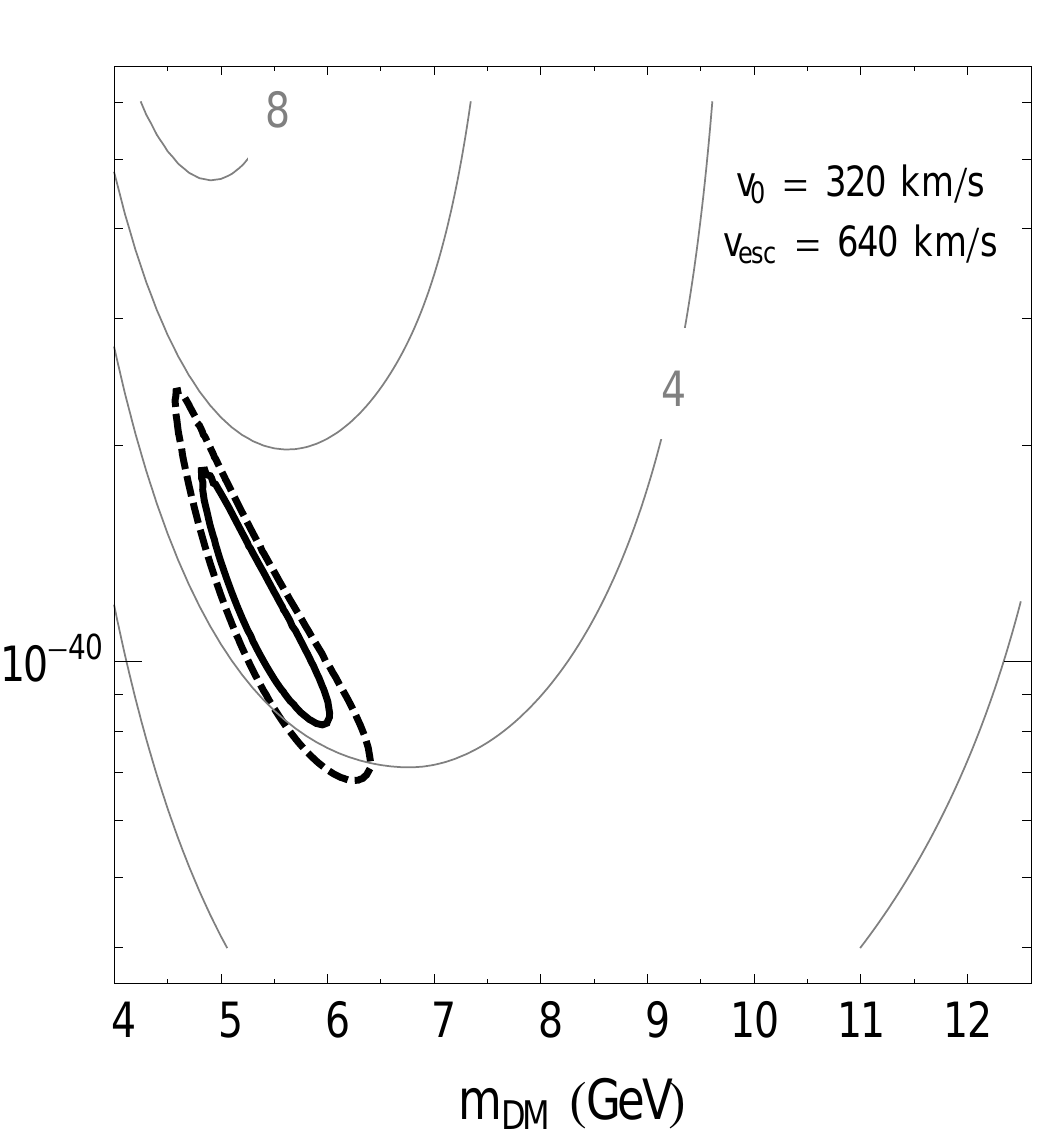}}
\caption{The 90\% (solid) and 99\% (dashed) confidence level contours for the spectrum of events observed by CoGeNT, for 9 choices of the velocity distribution parameters ($v_0$ and $v_{\rm esc}$). Also shown are contours for the predicted fractional modulation (given as a percentage of the overall rate) over the energy range of 0.5 to 3.0 keVee.}
\label{contours}
\end{figure*}

\begin{figure*}[t]
\centering
{\includegraphics[angle=0.0,width=2.49in]{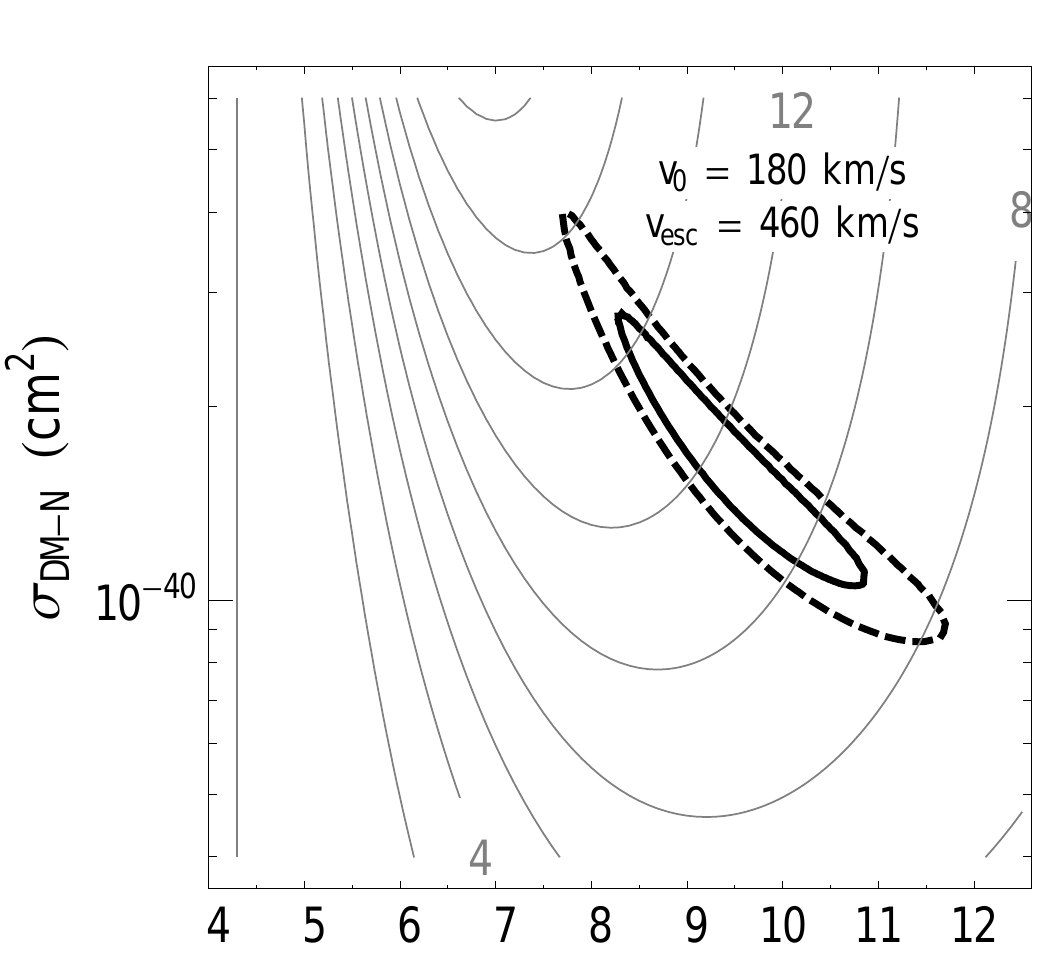}}
{\includegraphics[angle=0.0,width=2.24in]{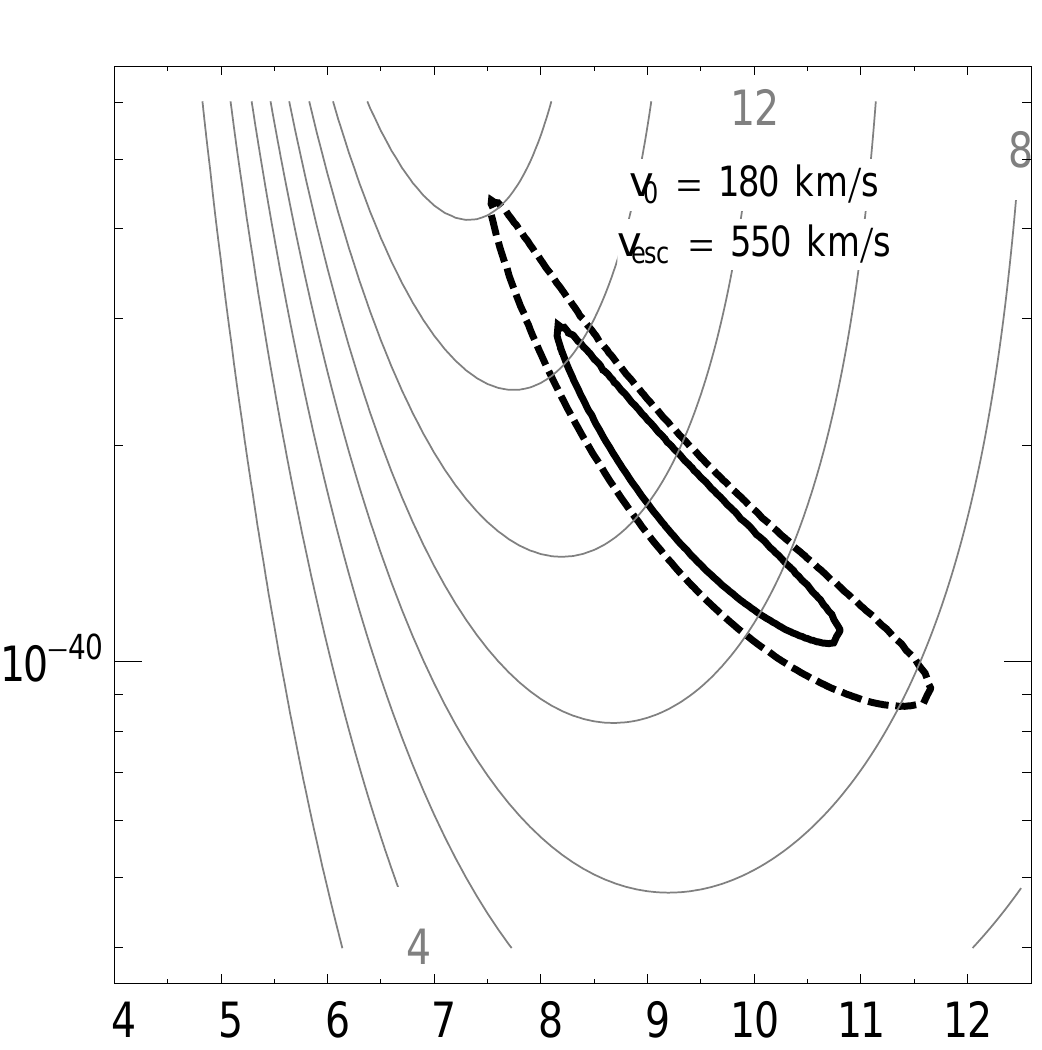}}
{\includegraphics[angle=0.0,width=2.24in]{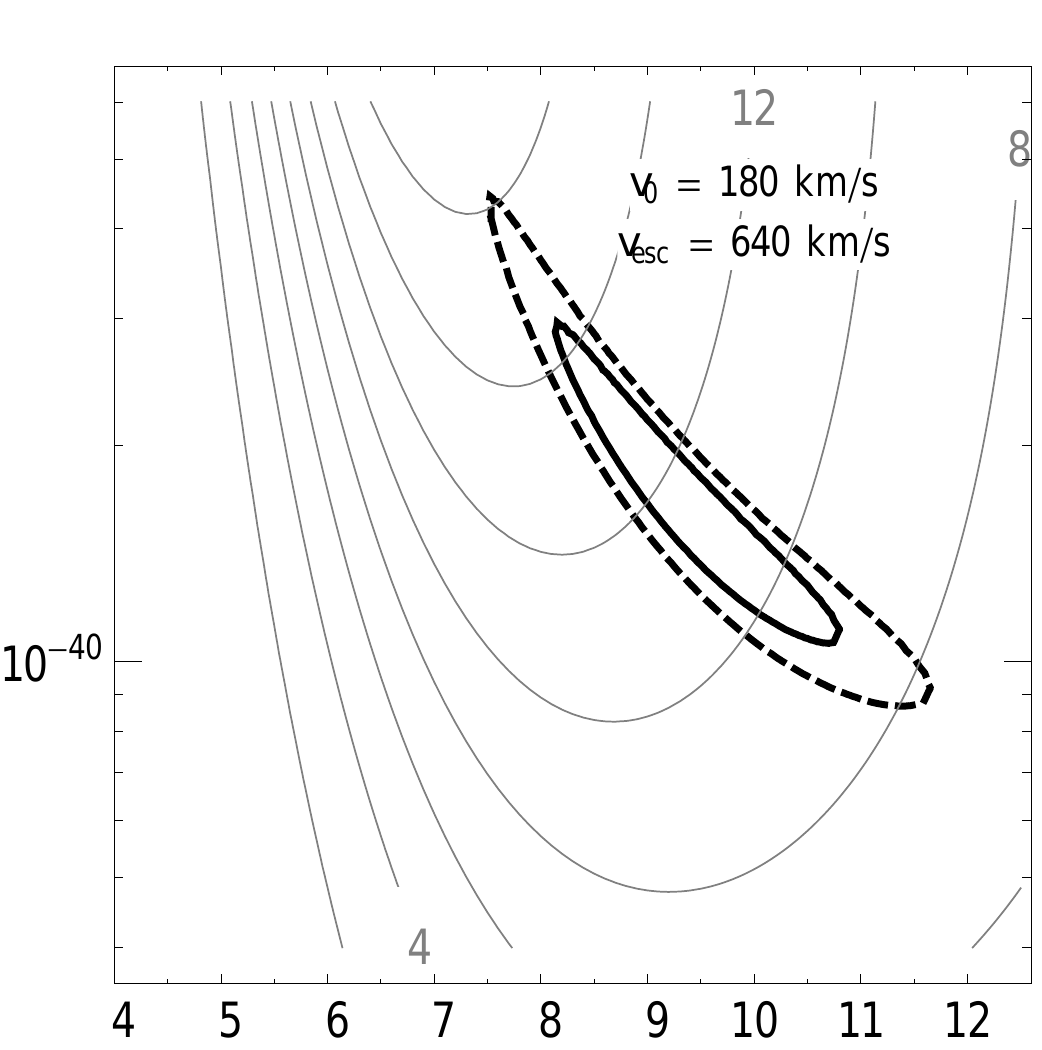}}\\
{\includegraphics[angle=0.0,width=2.49in]{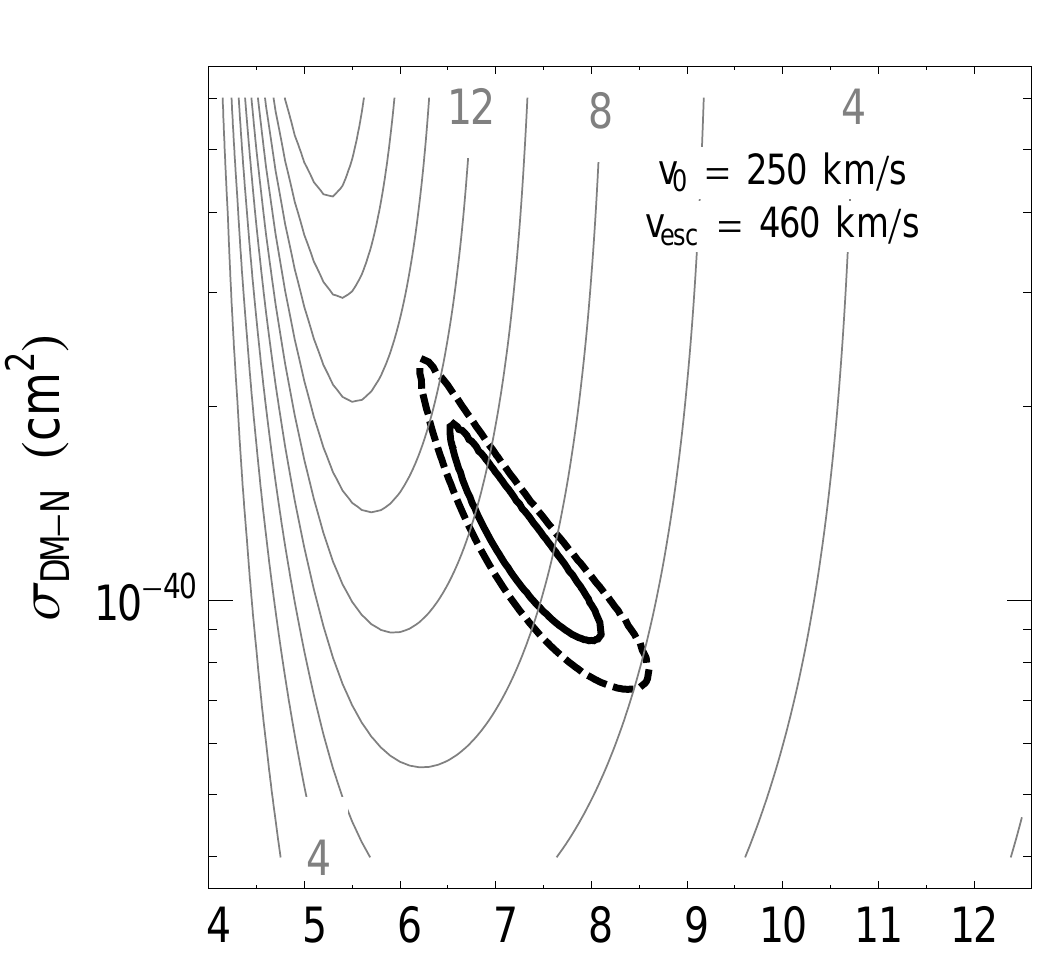}}
{\includegraphics[angle=0.0,width=2.24in]{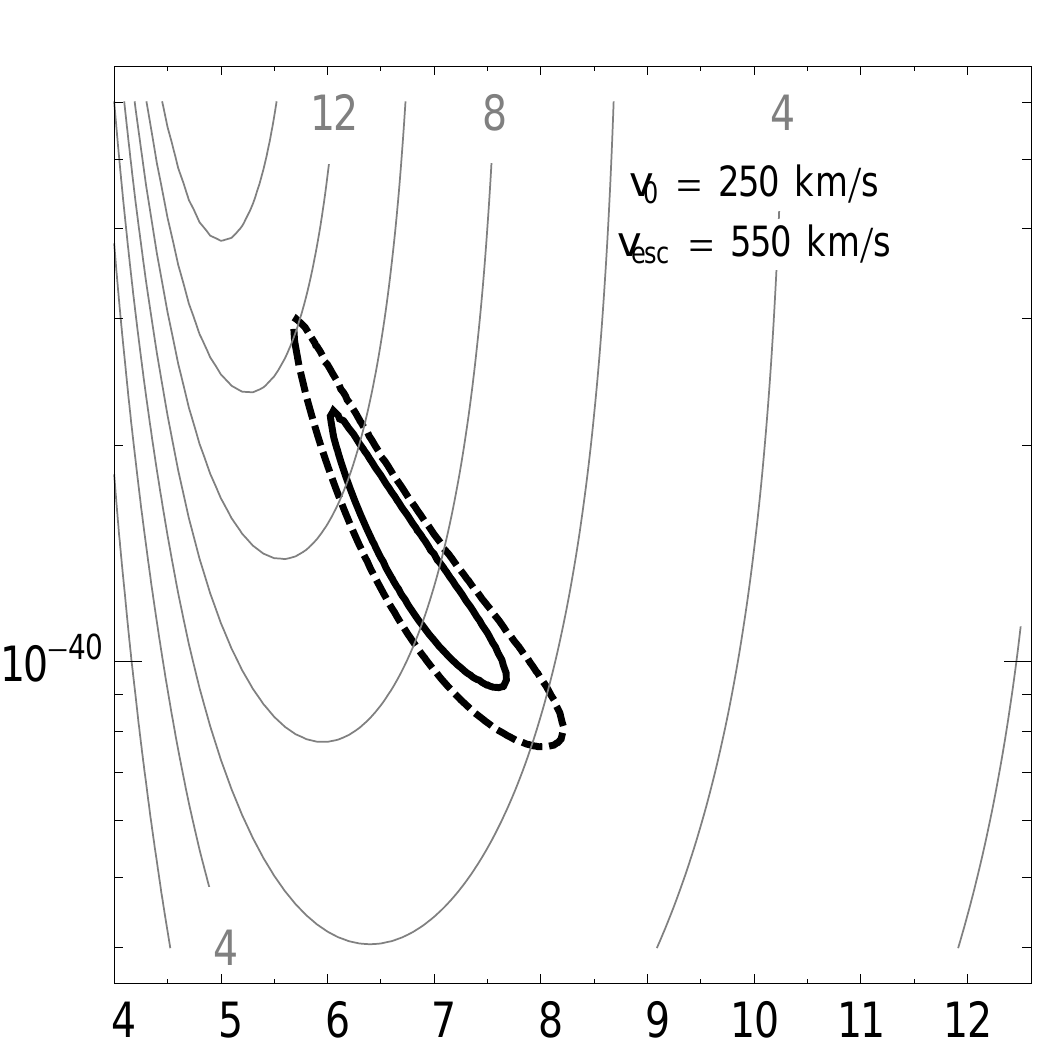}}
{\includegraphics[angle=0.0,width=2.24in]{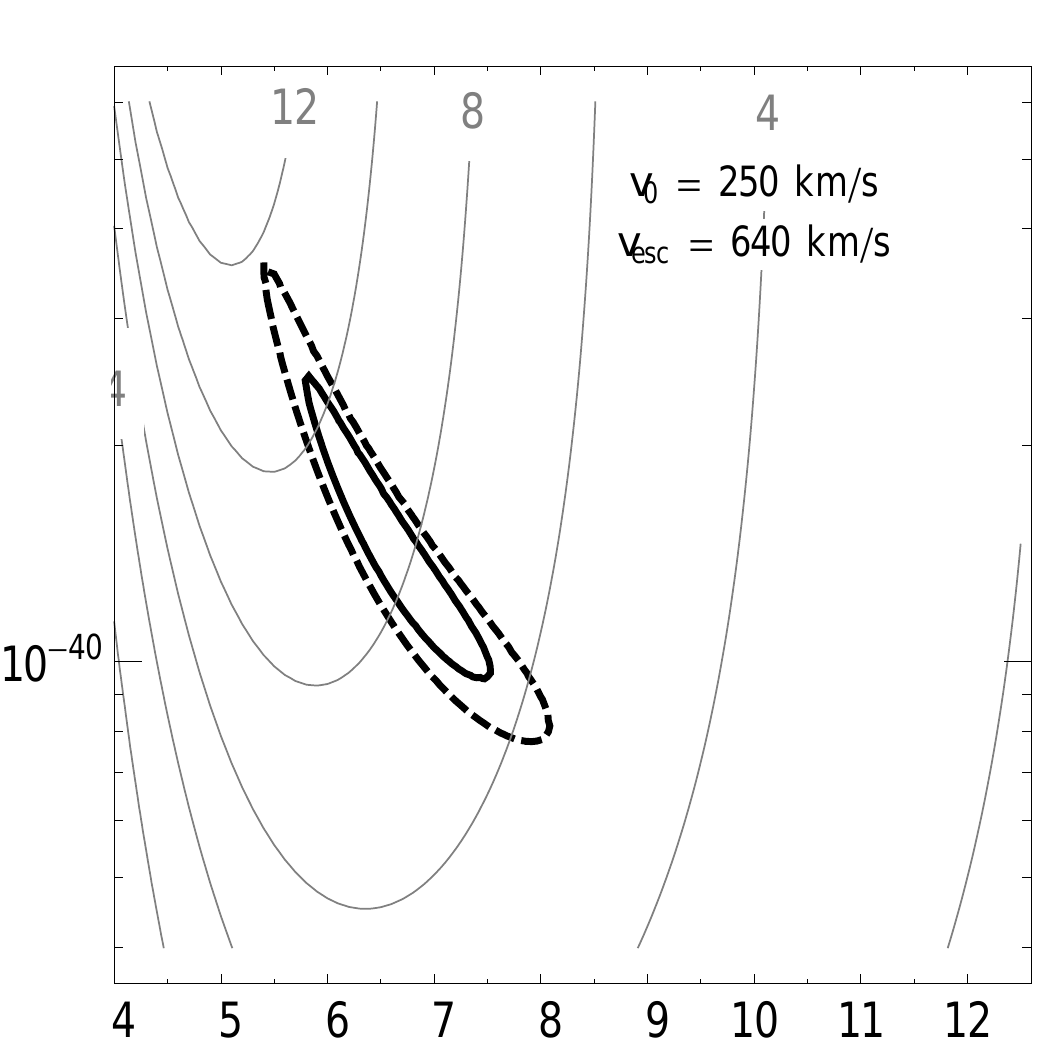}}\\
{\includegraphics[angle=0.0,width=2.49in]{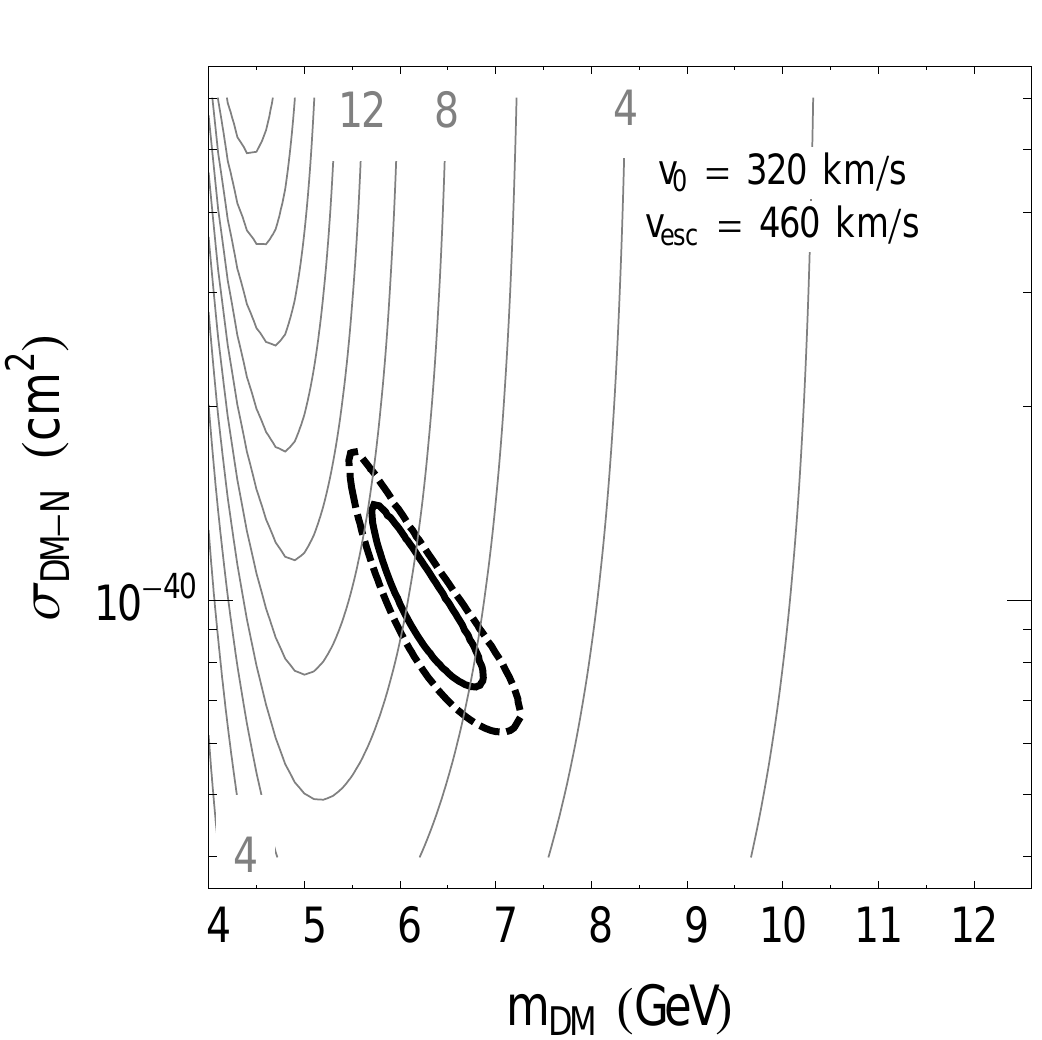}}
{\includegraphics[angle=0.0,width=2.24in]{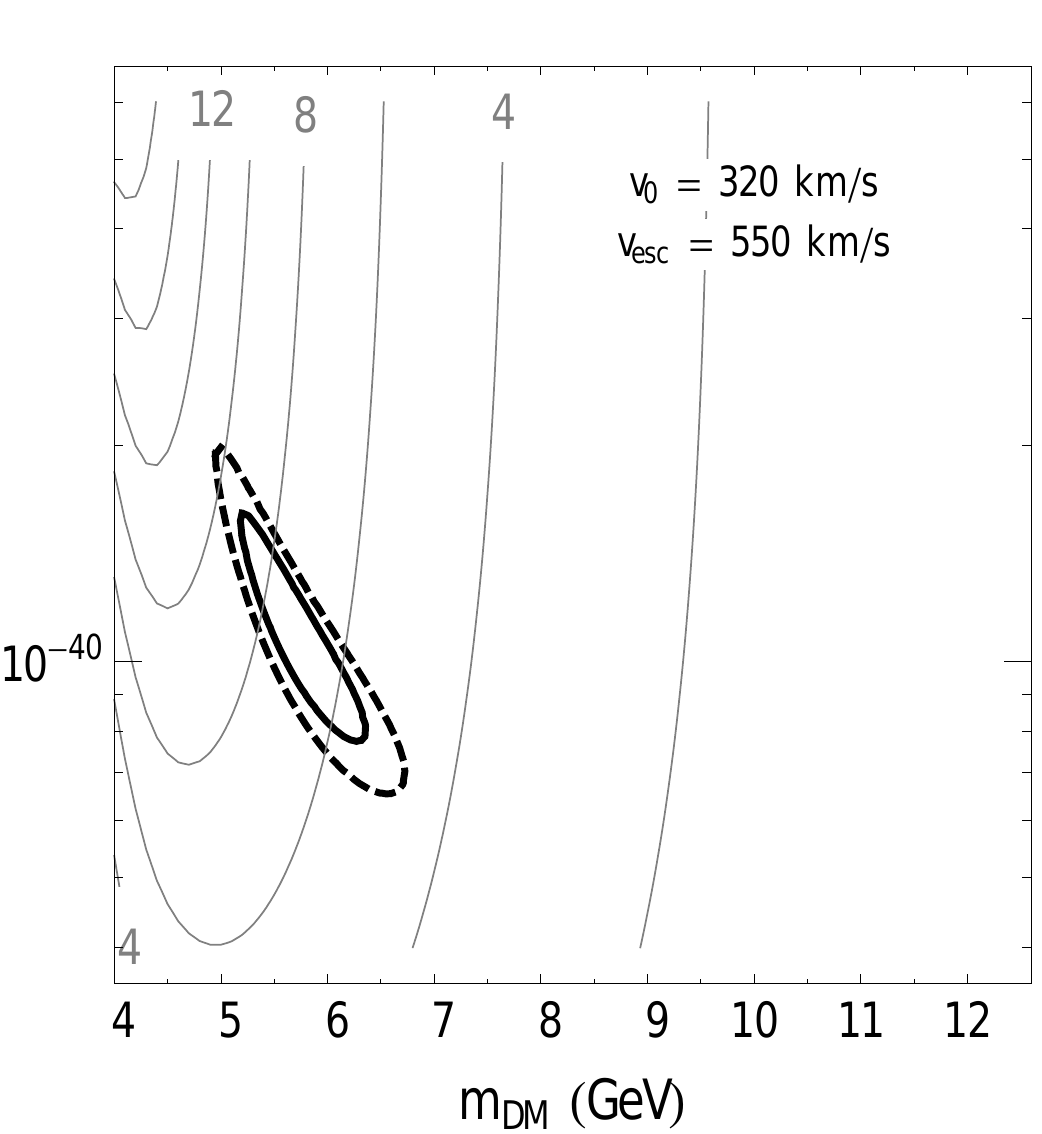}}
{\includegraphics[angle=0.0,width=2.24in]{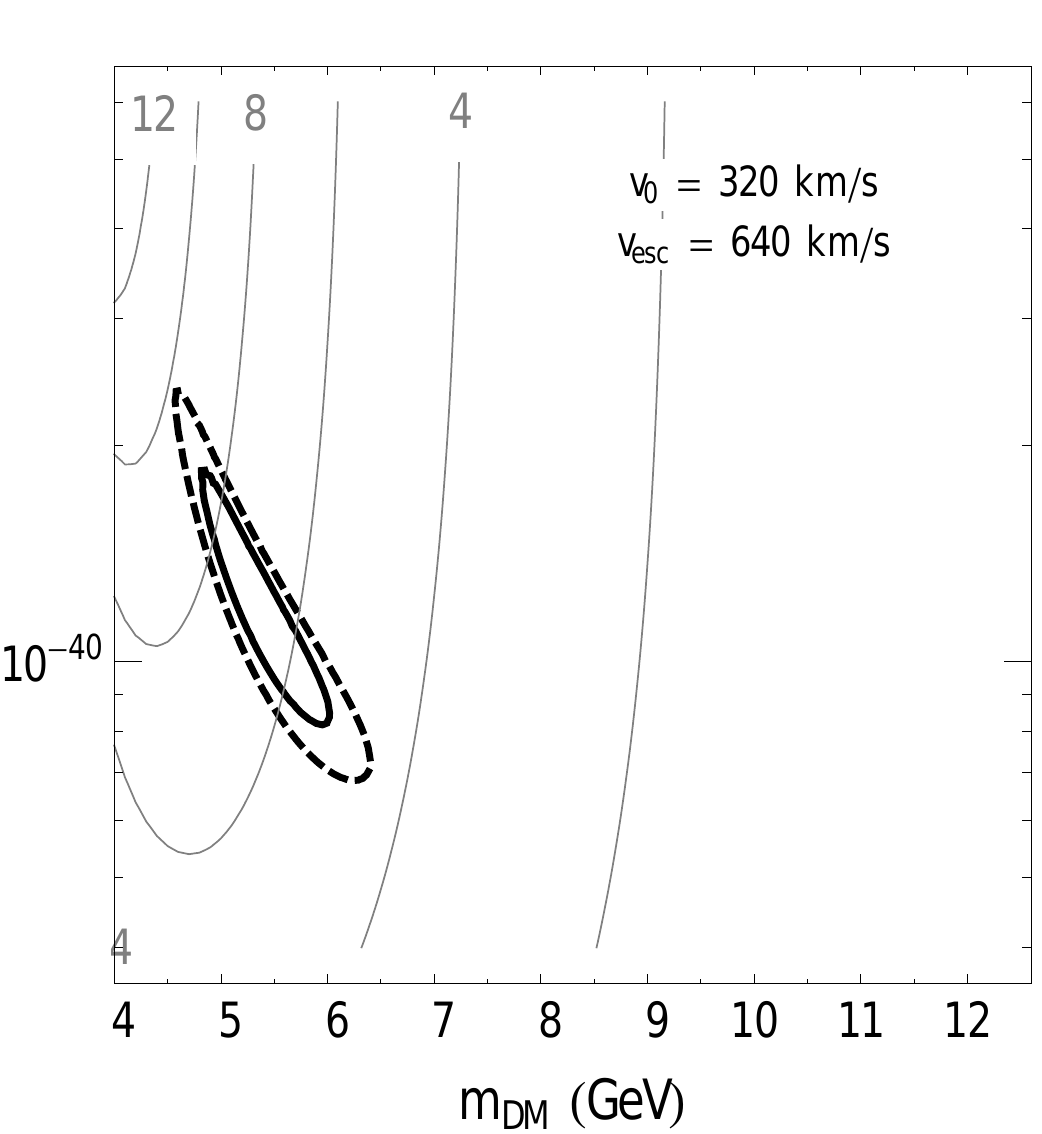}}
\caption{The same as Fig.~\ref{contours}, except that the contours for the predicted fractional modulation (given as a percentage of the overall rate) correspond to the energy range of 0.5 to 0.9 keVee.}
\label{contours2}
\end{figure*}

Considering a wider range of dark matter masses, cross sections and velocity distributions, we show in Figs.~\ref{contours} and \ref{contours2} the range of parameter space that provides a good fit to the spectrum observed by CoGeNT. We find that for a reasonable range of velocity distributions, the CoGeNT spectrum can be well fit by dark matter particles with masses in the range of approximately 4.5 to 12 GeV. Here, we have allowed the normalization of the flat background to float, but have removed the $L$-shell peaks according to the parameters listed in Table.~1. Also shown in these figures are contours which denote the fractional annual modulation (as a percentage of the average rate) that is predicted to be observed by CoGeNT over an energy range of 0.5-3.0 keVee (Fig.~\ref{contours}) and 0.5 to 0.9 keVee (Fig.~\ref{contours2}). In the next section, we will compare the predicted and observed annual modulations in more detail.

%\begin{twocolumn}
%\end{twocolumn}

%\begin{twocolumn}
%\end{twocolumn}

\section{CoGeNT's Annual Modulation}
\label{annual}

If the excess of events reported by CoGeNT is in fact the result of elastically scattering dark matter particles, then we should expect a degree of seasonal variation in the event rate. Due to the Earth's motion around the Sun, the rate of dark matter recoil events is predicted to vary throughout the year, peaking within several weeks of late May or early June~\cite{modulation}. In a previous study, we predicted that if CoGeNT's excess is the result of dark matter, the signal rate (not including backgrounds) should modulate at a level of between 1\% and 21\% over the energy range of 0.4 to 1.0 keVee (and between 1\% and 16\% over 0.5 to 0.9 keVee)~\cite{kelso}. This fractional modulation is defined as $(R_{\rm summer}-R_{\rm winter})/(2 R_{\rm ave})$, where $R_{\rm summer}$ and $R_{\rm winter}$ denote the maxima and minima of the rate. If non-modulating backgrounds are included, the predicted fractional modulation will be diluted accordingly.

From the contours shown in Figs.~\ref{contours} and \ref{contours2}, we see that the newly reported CoGeNT spectrum leads to an anticipated annual modulation at the level of 5\% to 16\% between 0.5 and 0.9 keVee and between 4\% to 10\% between 0.5 and 3.0 keVee. If the rate observed by CoGeNT did not demonstrate an annual modulation at approximately this magnitude, it would be difficult to interpret their excess events as a product of elastically scattering dark matter.

\begin{figure}[t]
\centering
{\includegraphics[angle=0.0,width=3.2in]{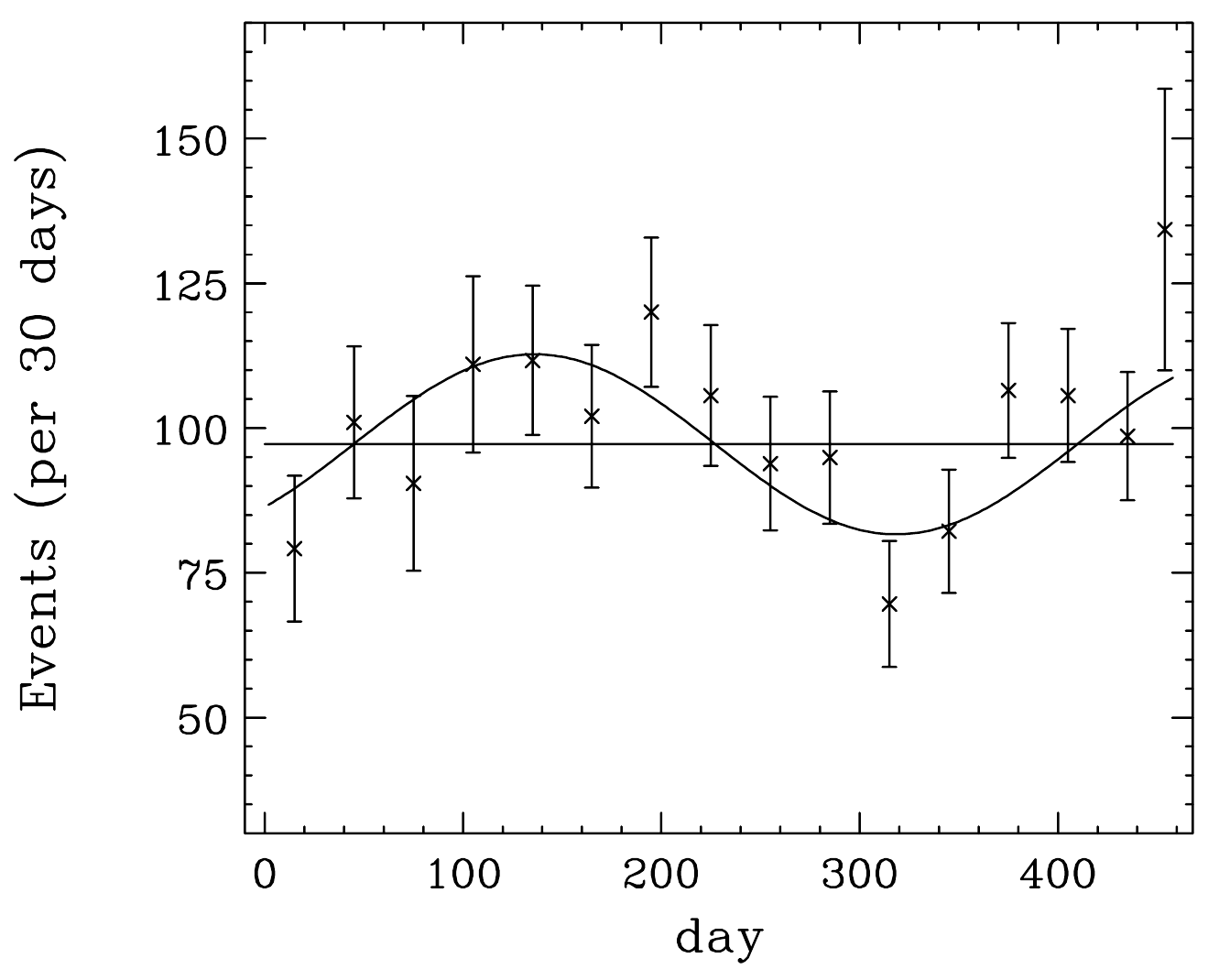}}
\caption{The rate of events between 0.5 and 3.2 keVee observed by CoGeNT, as a function of time, after the subtraction of L-shell peaks. Times are given in units of days since the beginning of CoGeNT's data taking (Dec.~4, 2009). The solid curve represents the best fit annual modulation (16\%, peaking at April 18), while the flat line is the constant rate with the best fit normalization.}
\label{mod}
\end{figure}

%\begin{twocolumn}
\begin{figure}[!]
\centering
{\includegraphics[angle=0.0,width=3.2in]{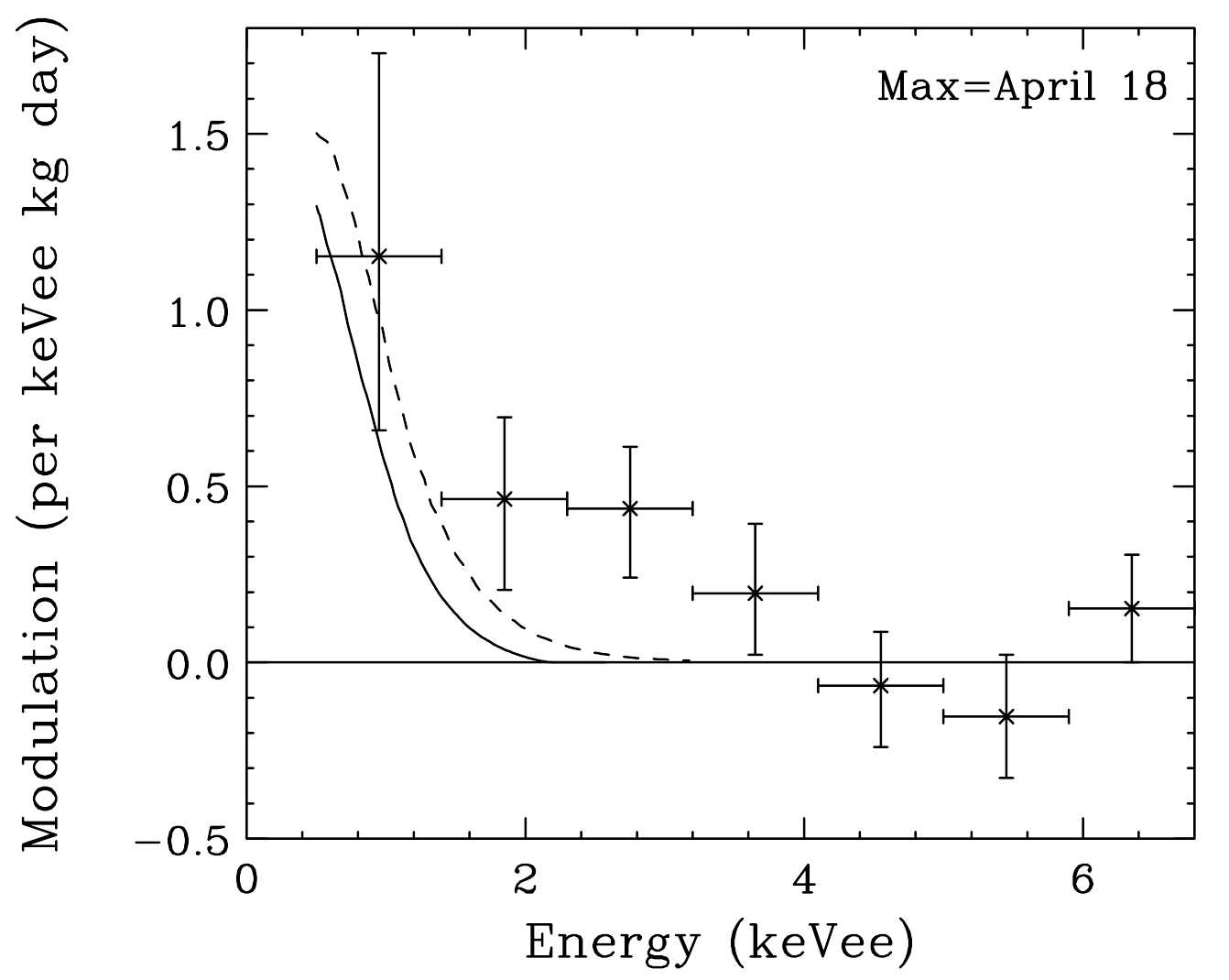}}
{\includegraphics[angle=0.0,width=3.2in]{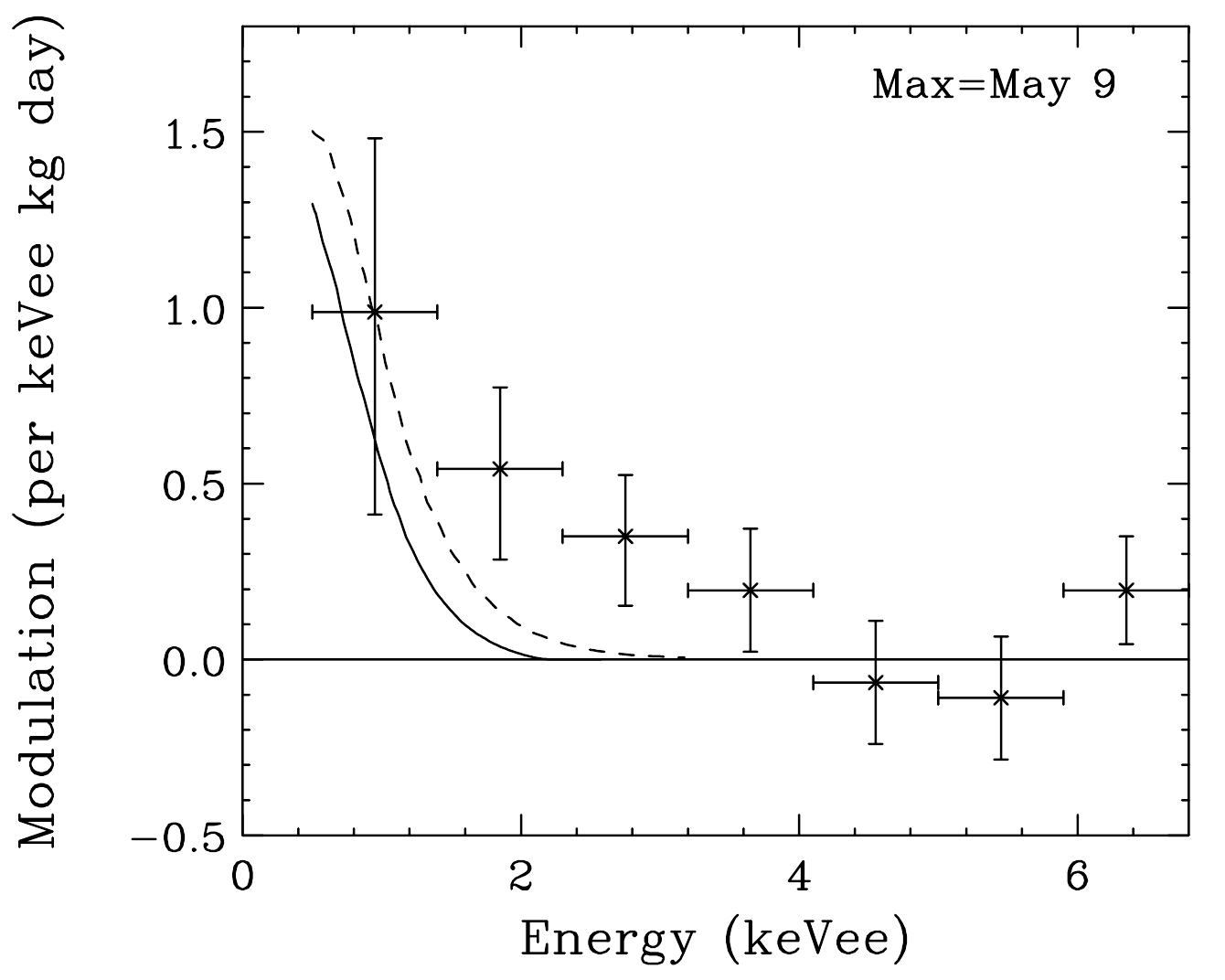}}
{\includegraphics[angle=0.0,width=3.2in]{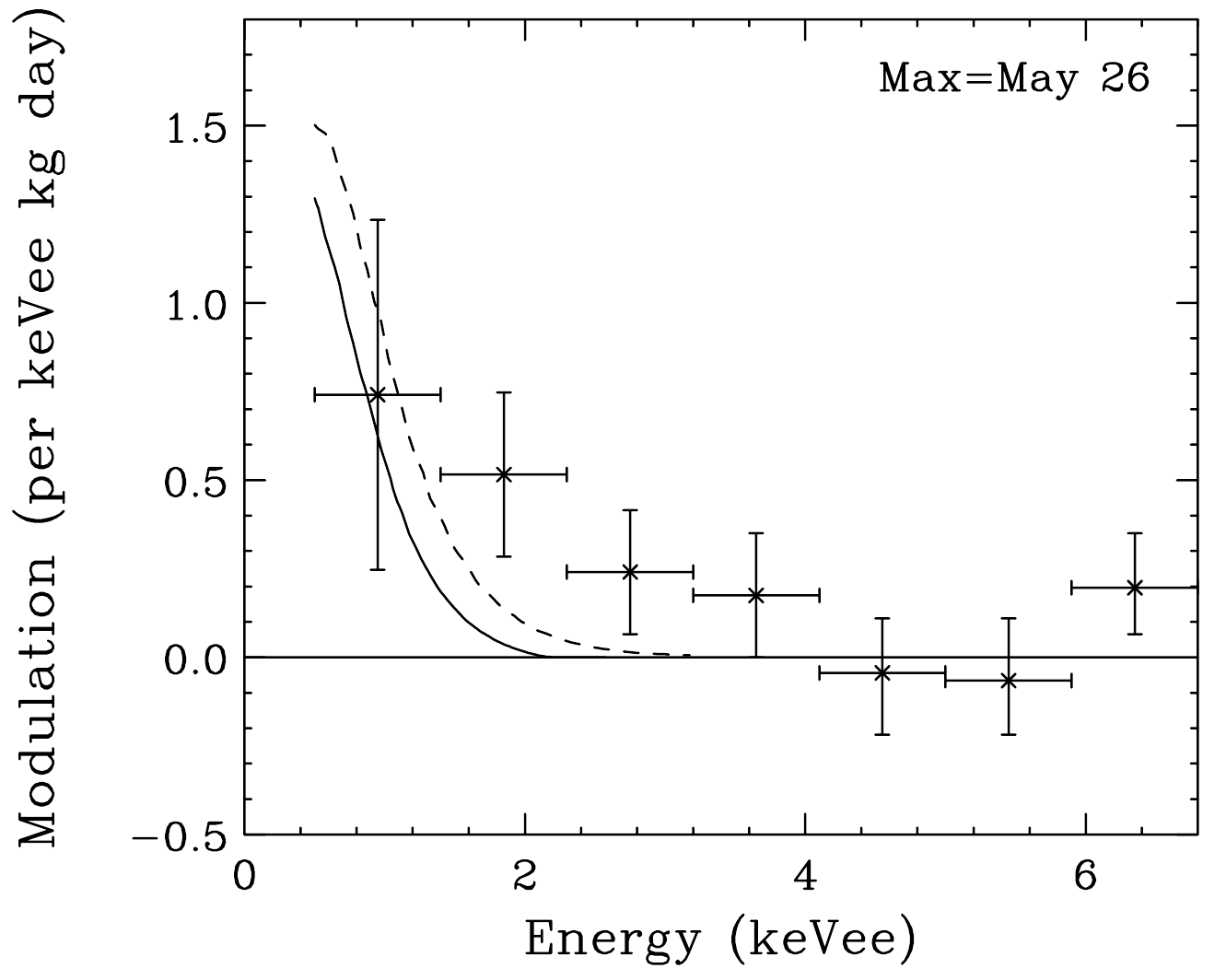}}
\caption{The spectrum of the annual modulation amplitude observed by CoGeNT for three choices of the phase. Also shown for comparison is the modulation spectrum predicted for two dark matter scenarios: $m_{\rm DM}=7$ GeV, $v_0=250$ km/s (solid) and $m_{\rm DM}=11$ GeV, $v_0=180$ km/s (dashed); each with $\sigma_{{\rm DM}-N}=1.2\times 10^{-40}$ cm$^2$ and $v_{\rm esc}=550$ km/s.}
\label{amp}
\end{figure}
%\end{twocolumn}

In Fig.~\ref{mod}, we plot the rate of events observed by CoGeNT with energies between 0.5 and 3.2 keVee as a function of time, after subtracting the contribution from L-shell peaks. Based on our analysis of this data, we find that the presence of an annual modulation is favored over a flat event rate at a confidence level corresponding to 2.7$\sigma$ (the CoGeNT collaboration, in their own analysis, finds a similar significance of 2.8$\sigma$ for events between 0.5 and 3.0 keVee~\cite{newcogent}). In particular, we find a modulation of 16$\pm$5\% (including the flat background, but after the subtraction of L-shell peaks), and with a phase that peaks at April 18$\pm$16 days. Again, the CoGeNT collaboration's analysis yields very similar conclusions (16.6$\pm$3.8\%, peaking at April 16$\pm$12 days).

Comparing the phase of this observed modulation to that reported by the DAMA/LIBRA collaboration, we find that both CoGeNT and DAMA/LIBRA prefer a peak rate that occurs somewhat earlier than the late May/early June region typically expected from dark matter (the phase of DAMA/LIBRA modulation between 2 and 4 keV and between 2 and 6 keV, has been reported as May 16$\pm$7 days and  May 26$\pm$7 days, respectively). The combination of CoGeNT and DAMA data collectively favor a modulation that peaks in early May. Studies based on N-body simulations find that 68\% of all realizations feature a peak rate that is within $\pm 20$ days from late May/early June~\cite{Kuhlen:2009vh}.  Thus we conclude that the phase of the modulation favored by CoGeNT is consistent with that reported by DAMA/LIBRA, and with that expected from elastically scattering dark matter.

Although the statistics provided by CoGeNT are limited, we can begin to study the spectrum of the observed modulation amplitude. In Fig.~\ref{amp}, we show the observed modulation amplitude, for three choices of the phase (peaking at April 18, May 9, and May 26). We find the presence of modulation in each of the three energy bins below 3.2 keVee, but no statistically significant modulation at higher energies. We also show in each of these frames the modulation spectrum that is predicted for two dark matter scenarios: $m_{\rm DM}=7$ GeV, $v_0=250$ km/s (solid) and $m_{\rm DM}=11$ GeV, $v_0=180$ km/s (dashed); each with $\sigma_{{\rm DM}-N}=1.2\times 10^{-40}$ cm$^2$ and $v_{\rm esc}=550$ km/s. At this point, we note that there appears to be somewhat more modulation observed at 1.4-3.2 keVee than is predicted, although more data will be needed to evaluate this issue with satisfactory statistical significance. The modulation in this energy range could be enhanced if the dark matter's velocity distribution were to depart significantly from the Maxwellian form that we have assumed.

\section{Results Of and Prospects For Other Direct Detection Experiments}
\label{others}

In this section, we discuss the implications of the results of other direct detection experiments on a dark matter interpretation of the CoGeNT spectrum and modulation. We will also discuss the prospects for other direct detection experiments which may be sensitive to dark matter in the $\sim$5-10 GeV mass range.

\subsection{Comparison With Results From DAMA/LIBRA}

The only direct detection experiment other than CoGeNT to report the observation of an annual modulation is DAMA/LIBRA~\cite{damanew}. The statistical significance of DAMA's modulation is very high (8.9$\sigma$), and demonstrates a phase which is compatible with that measured by CoGeNT (peaking at May 16$\pm$7 days between 2 and 4 keV and May 26$\pm$7 days between 2 and 6 keV, compared to April 18$\pm$16 days for CoGeNT). The combination of CoGeNT and DAMA/LIBRA data favor a modulation that peaks in early May, which is consistent with expectations from dark matter simulations~\cite{Kuhlen:2009vh}. 

\begin{figure}[t]
\centering
{\includegraphics[angle=0.0,width=3.2in]{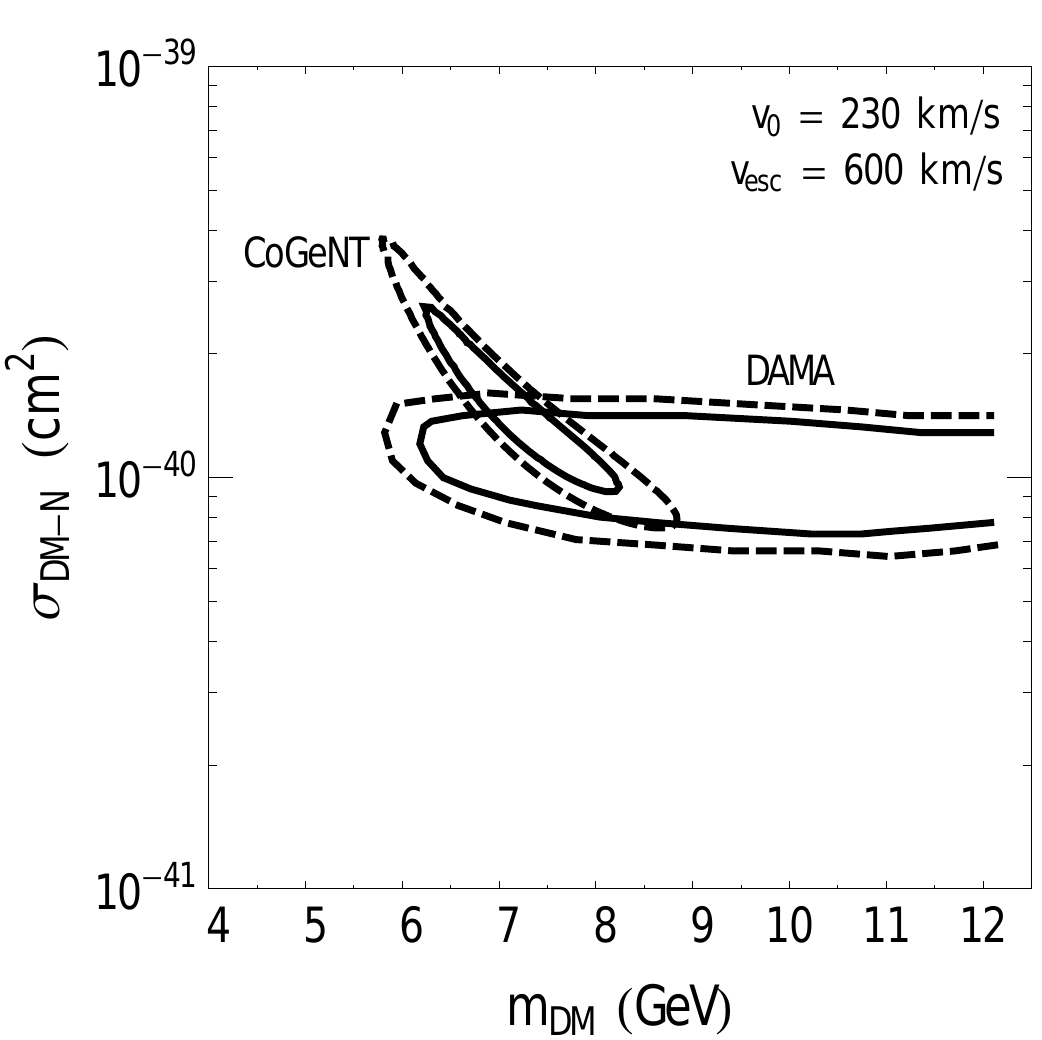}}
\caption{A comparison of the parameter space favored by the CoGeNT spectrum with that favored by the modulation spectrum reported by DAMA/LIBRA~\cite{consistent}. Good agreement is found, but somewhat large quenching factors for low energy nuclear recoils on sodium are required ($Q_{\rm Na}\sim 0.40-0.45$)~\cite{consistent}.}
\label{damafig}
\end{figure}

In Fig.~\ref{damafig}, we compare the regions of the dark matter parameter space favored by the CoGeNT spectrum to those favored by the modulation spectrum reported by DAMA/LIBRA (the DAMA region has been taken from Ref.~\cite{consistent}, and we have used the velocity distribution parameters from that study for comparison). The agreement is clearly very good, but requires the quenching factors for low energy nuclear recoils on sodium to be somewhat larger than are often assumed ($Q_{\rm Na}\sim 0.40-0.45$ rather than $Q_{\rm Na}\sim 0.3$, see also Ref.~\cite{quth})~\cite{consistent}. We have not included any effects of channeling~\cite{Bozorgnia:2010xy} in these results. If significant channeling occurs in DAMA's NaI crystals, the favored range of masses and cross sections would be modified.

\subsection{Constraints From CDMS and XENON}

\begin{figure}[t]
\centering
{\includegraphics[angle=0.0,width=3.2in]{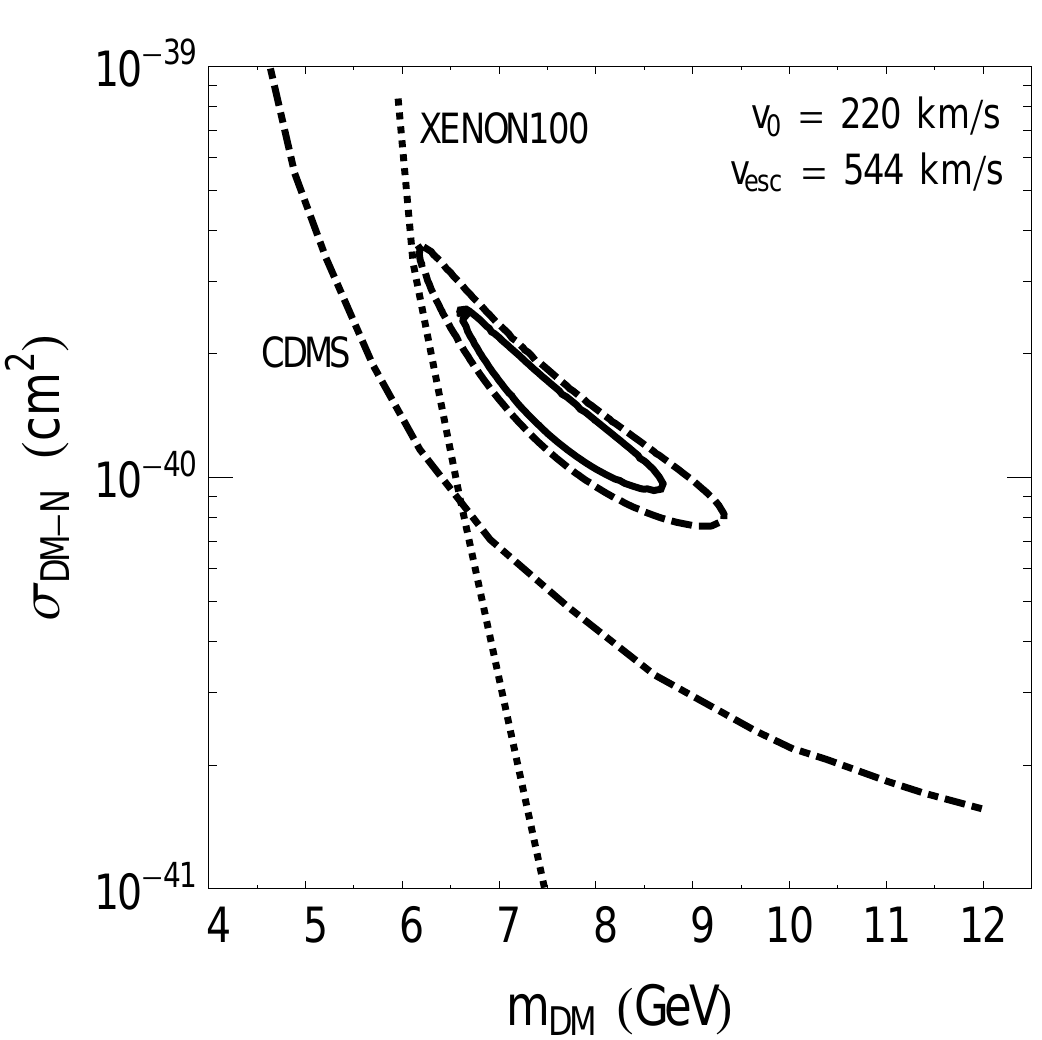}}
\caption{Constraints on light dark matter particles as presented by the CDMS (dot-dash)~\cite{Ahmed:2010wy} and XENON100 (dotted)~\cite{xenon100} collaboration. For a discussion of these constraints and their implications for CoGeNT, see the text and Refs.~\cite{Collar:2010ht,juan2,Collar:2011kf}.}
\label{excl}
\end{figure}

The CDMS and XENON100 collaborations have each presented results which they interpret to exclude or strongly constrain dark matter interpretations of the CoGeNT signal (see Fig.~\ref{excl}). Here, we will briefly review these results and discuss means by which they could potentially be reconciled with CoGeNT.

In April of 2011, the XENON100 collaboration presented the result of their first 100 live days of data~\cite{xenon100}, and conclude that (for a velocity distribution given by $v_0=220$ km/s, $v_{\rm esc}=544^{+64}_{-46}$ km/s) a dark matter particle with a mass of 7 GeV is required to possess a nucleon-level cross section less than $\sim$$3\times 10^{-41}$ cm$^2$. The constraint falls off quickly with the mass of the dark matter, however; for a 6 GeV mass, for example, the quoted constraint is weaker by a factor of five, to $\sim$$1.5\times 10^{-40}$ cm$^2$ (see Fig.~\ref{excl}). At face value, this result appears to exclude the region of parameter space consistent with the spectrum reported by CoGeNT. There are a number of ways, however, in which this constraint could be significantly weaker than it appears. Firstly, any uncertainties in the scintillation efficiency of liquid xenon, $L_{\rm eff}$, and/or in the quenching factor of germanium, could impact the corresponding constraints for dark matter particles with mass in the range of interest. The XENON100 constraints have been derived using measurements of $L_{\rm eff}$ as described in Refs.~\cite{plante,manzur}, which have been criticized in Refs.~\cite{Collar:2010ht,juan2}. Even modest changes to these values at the lowest measured energies ($\sim$3-4 keV) can lead to much weaker constraints on light dark matter particles. It has been been argued that the relatively large (9.3 eV) band-gap in liquid xenon should lead to suppression of xenon's sensitivity to nuclear recoils in the energy range of interest (see Ref.~\cite{Collar:2010ht} and references therein). Many of these issues also apply to constraints on light dark matter making use of only the ionization signal in liquid xenon detectors~\cite{Angle:2011th}.

Alternatively, any apparent conflict between CoGeNT and XENON100 could be resolved if dark matter particles couple differently to protons and neutrons~\cite{zurek,feng}. In particular, for a ratio of these couplings given by $f_n/f_p \approx -0.7$, the constraint from xenon-based experiments is weakened by a factor of $\sim$20 relative to that found in the $f_n=f_p$ case. 

The rate of low energy events reported by the CDMS collaboration is also somewhat lower than those observed by CoGeNT~\cite{Ahmed:2010wy,Akerib:2010pv}. The degree to which these spectra are discrepant has been discussed elsewhere, and depends on issues such as the precise calibration of the CDMS energy scale, and on the choice of basing the CDMS measurement on the single detector with the lowest rate, or on the average of the eight detectors (see Ref.~\cite{Collar:2011kf} and the appendix of Ref.~\cite{Ahmed:2010wy} for opposing viewpoints on this and related issues).

As both CDMS and CoGeNT use germanium (along with silicon in the case of CDMS) as their dark matter target, differences in their relative rates cannot be accounted for by varying the ratio of $f_n$ and $f_p$. One possibility is that the relatively warm temperature of CoGeNT compared to CDMS ($T\approx90$ K vs. 0.040 K) leads to a fraction of events to be channeled at CoGeNT, but not at CDMS. Although theoretical estimates suggest that the probability of channeling is too low to account for this discrepancy~\cite{Bozorgnia:2010ax}, the non-occurrence of channeling in germanium crystals is yet to be experimentally confirmed.

\subsection{Predictions For COUPP and CRESST}

%\begin{twocolumn}
\begin{figure*}[t]
\centering
{\includegraphics[angle=0.0,width=3.4in]{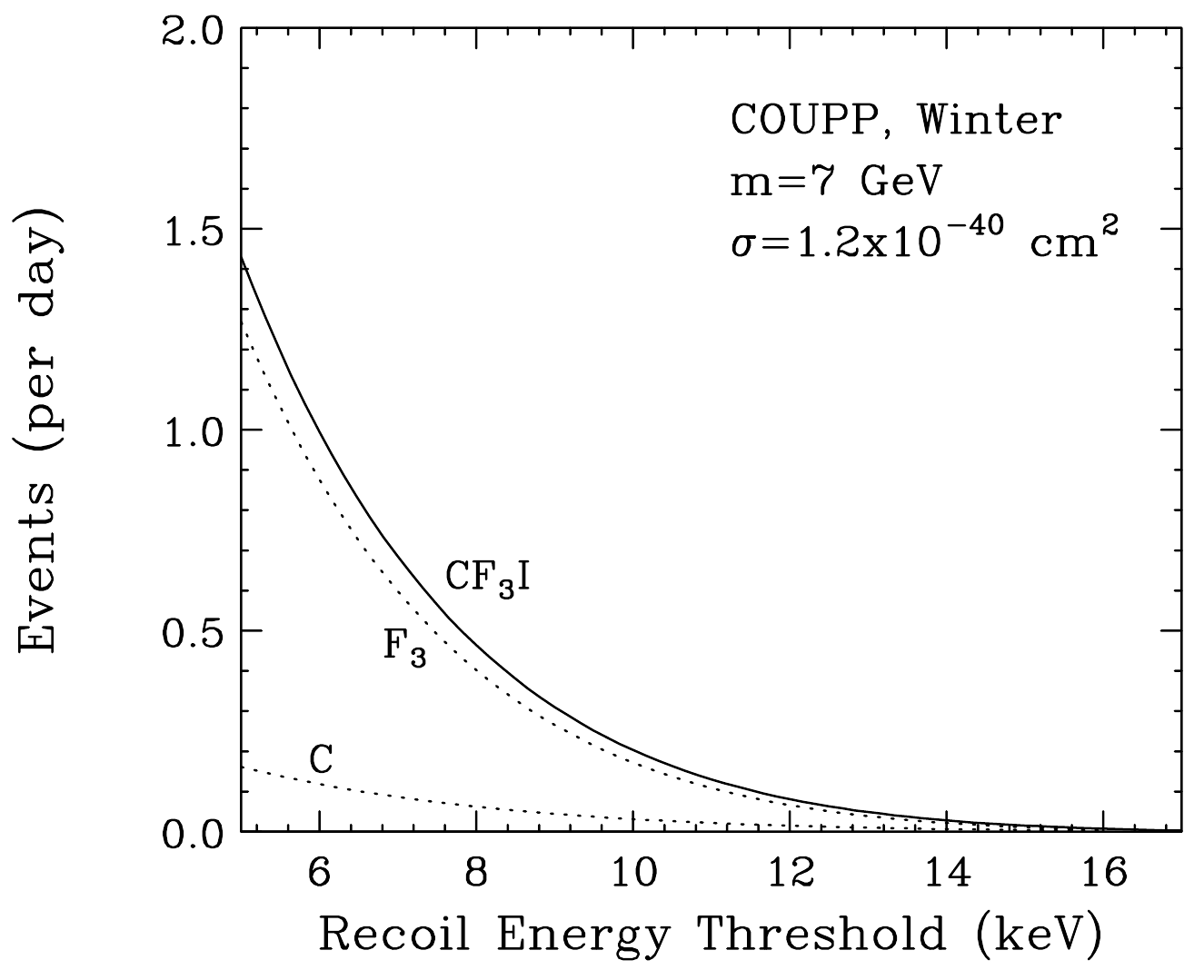}}
{\includegraphics[angle=0.0,width=3.4in]{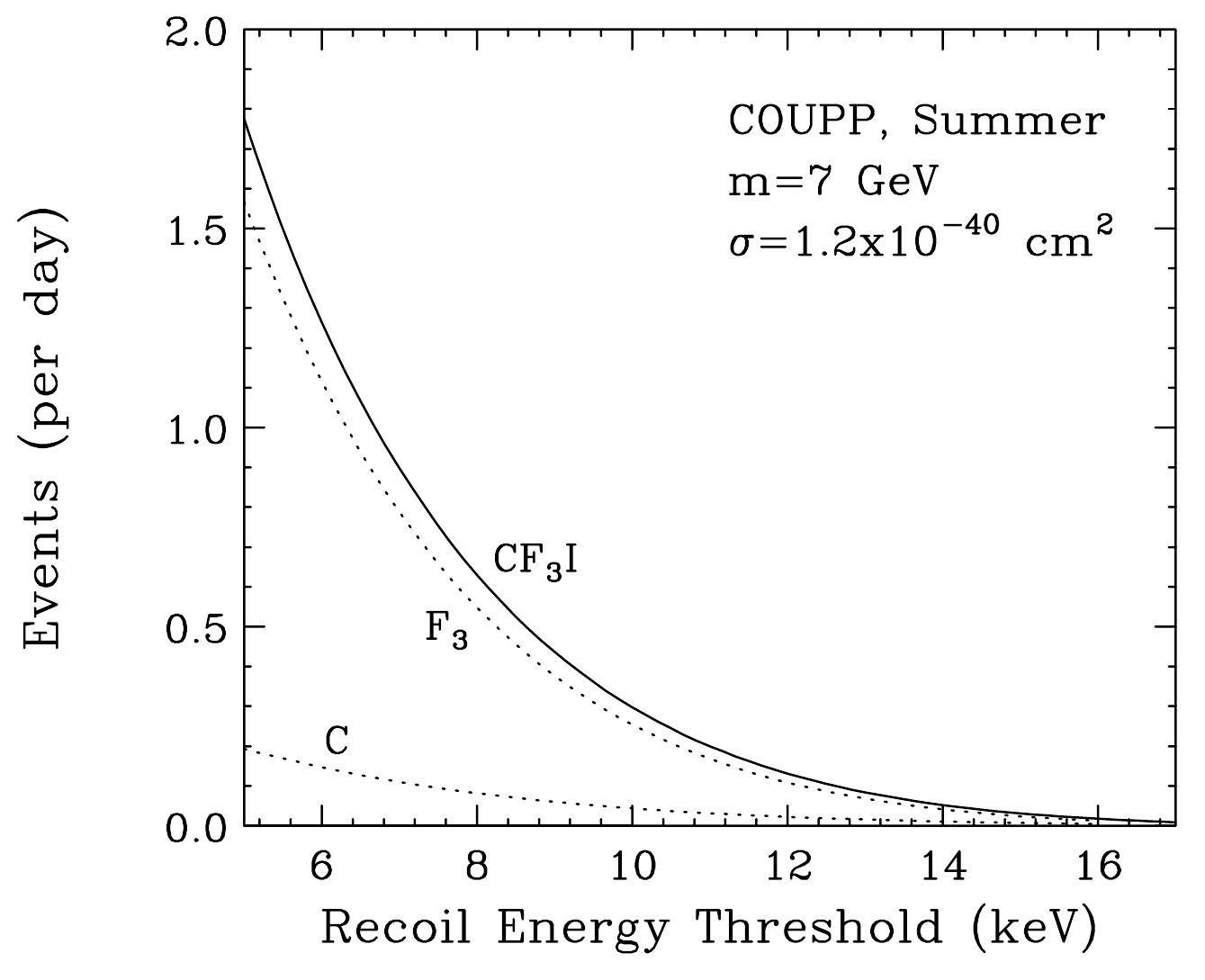}}
\caption{The event rate (with a fidicual mass of 3.3 kg) at COUPP from a CoGeNT-like dark matter particle as a function of the recoil energy threshold and in the winter (left) and summer (right). The overall rate is predicted to vary dramatically with threshold.}
\label{coupp}
\end{figure*}
%\end{twocolumn}

Experiments which make use of relatively light elements, such as CRESST (CaWO$_4$) and COUPP (CF$_3$I) are potentially well suited to detect and study light dark matter particles. If CoGeNT is in fact observing the elastic scattering of dark matter particles, these experiments should also be capable of observing such events.

In a number of conference talks given over the past year, members of the CRESST collaboration have reported an excess of events which appears to be consistent with the elastic scattering of dark matter~\cite{cresst}. More specifically, based on approximatley 700 kg-days of data, CRESST observes a rate of events in their oxygen band (events which are consistent with the recoil of an oxygen nucleus) which is in excess of their expected backgrounds at the level of 4.6$\sigma$. For dark matter particles with a mass in the range favored by CoGeNT, spin-independent scattering is expected to occur mostly with oxygen nuclei (rather than with CRESST's tungsten or calcium). Although more details of this analysis will be needed before any firm conclusions can be drawn, the preliminary results from CRESST appear to favor dark matter particles with a mass and elastic scattering rate similar to that implied by CoGeNT~\cite{cresst,consistent}. In particular, it was recently reported that the spectrum of CRESST's events is best fit by a dark matter particle with a mass of 13 GeV and a cross section with nucleons of $3 \times 10^{-41}$ cm$^2$, although the confidence contours around this best fit model have not been reported~\cite{cresst}. We eagerly await further details pertaining to the CRESST analysis.

The COUPP collaboration has very recently begun operation of their 4 kg chamber at SNOLAB (3.3 kg fiducial). A dark matter particle near the center of the region preferred by CoGeNT and DAMA/LIBRA is predicted to generate $\sim$0.7 events at COUPP per day, when running with a recoil energy threshold of $\sim$7 keV. If their backgrounds are as low as anticipated, COUPP could rapidly accumulate a significant excess of events.

Over the past several months, the 4 kg COUPP chamber has been operated at temperatures and pressures corresponding to three different recoil energy thresholds, estimated at 7, 10 and 15 keV. In Fig.~\ref{coupp}, we show the event rate predicted at COUPP for a CoGeNT-like dark matter particle, as a function of the recoil energy threshold. From this figure, it is clear that this variation of threshold is predicted to result in a dramatic variation in the rate of dark matter induced events. The approximately 20 live days of data taken at each of 7 and 10 keV thresholds~\cite{coupptalk} should be anticipated to contain $\sim$14 and $\sim$4 events, respectively, from dark matter scattering. In contrast, less than one event per month is anticipated when running with a threshold of 15 keV. With a sufficiently large exposure, it may also be possible for COUPP to observe season variations in their event rate. The predicted rate with a 15 keV threshold, in particular, can vary by a factor of 2-3 between summer and winter.

Both COUPP and CRESST could enhance their event rate from light dark matter particles per target mass by adopting a target material which does not contain heavy nuclei. The COUPP collaboration, for example, has considered replacing their CF$_3$I target with C$_3$F$_8$, resulting in a roughly 50\% higher event rate from light dark matter particles (after accounting for the lower density). Similarly, the CRESST collaboration is considering using Al$_2$O$_3$ as a target more favorable for low mass dark matter~\cite{cresst}.

%%%%%%%%%%%%%%

\section{Summary and Conclusions}
\label{summary}

The possibility that the excess of low energy events as originally reported by the CoGeNT collaboration last year~\cite{cogent} is the result of elastically scattering dark matter particles has received a great deal of attention~\cite{attention,kelso,consistent,Lisanti:2010qx,zurek,feng}. The most clear and straightforward test of this hypothesis was to observe whether or not the rate of this excess modulated with time, and if so whether its modulation was consistent in amplitude, phase, and period with the annual modulation predicted for elastically scattering dark matter~\cite{modulation,consistent,kelso,integrating}. With the most recent results from CoGeNT, based on 15 months of data taking, we have learned that such a modulation does in fact appear to be present (with a statistical significance of 2.7-2.8$\sigma$), and is consistent with a simple interpretation as a relatively light dark matter particle ($m_{\rm}\approx$5-12 GeV) with a sizeable elastic scattering cross section with nucleons ($\sigma_{{\rm DM}-N} \sim 10^{-40}$ cm$^2$).

In this paper, we have independently analyzed the CoGeNT data (as made available by the CoGeNT collaboration) and reached similar conclusions to those presented by the CoGeNT collaboration~\cite{newcogent}. In particular, we find that over the energy range of 0.5 to 3.2 keVee, the overall rate (after the substraction of L-shell peaks) modulates with an amplitude of $16\pm5\%$, with a period consistent with one year, and with a phase that peaks at April $18\pm16$ days. If the true phase peaks in early May, this would represent a modulation consistent with that reported by the DAMA/LIBRA collaboration~\cite{damanew}.

Looking forward, it is clear that more data will be required to better explore the dark matter interpretation of the CoGeNT signal. Although the current data set is sufficient to identify a modestly statistically signficant annual modulation, the energy spectrum of this modulation can not yet be studied in much detail. If the existing CoGeNT detector is able to recommence its operation following the recent fire in the Soudan Mine, the additional exposure will certainly be valuable in further efforts to the characterize the signal in question. Furthermore, the first of four detectors to make up the CoGeNT-4 (C4) experiment is planned to be deployed later this year. Each of these four detectors will offer a fiducial mass two to three times larger than in the current CoGeNT detector. Approximately 17 months of data taken with the first of the C4 detectors is projected to identify the presence of annual modulation with a significance of 5$\sigma$. The entire C4 experiment will offer the ability to measure the spectrum of this modulation in considerable detail, allowing us to begin to disentangle the mass and cross section of the dark matter particle from the dark matter's velocity distribution.

If the excess CoGeNT events and their modulation is the result of an elastically scattering dark matter particle, then the CRESST and COUPP experiments should also be capable of observing a signficant rate of dark matter induced events. The CRESST collaboration has reported the observation of an excess of events roughly consistent with that anticipated from a CoGeNT-like dark matter particle. We eagerly await further details from CRESST, and the presentation of the first results from COUPP at SNOLAB.

\section*{Acknowledgements} We would like to thank the CoGeNT collaboration for volunteering to make its data available for public analysis. We would also like to thank Juan Collar for many helpful discussions. This work has been supported by the US Department of Energy, including grant DE-FG02-95ER40896, and by NASA grant NAG5-10842.

\end{document}